\documentclass[conference]{IEEEtran}
\usepackage{ifpdf}

\usepackage{cite}

\ifCLASSINFOpdf
   \usepackage[pdftex]{graphicx}
\usepackage{epstopdf}
\else
 \usepackage[dvips]{graphicx}
  \DeclareGraphicsExtensions{.eps}
\fi
\usepackage[cmex10]{amsmath}
\usepackage{amssymb}
\usepackage{algorithm}
\usepackage{stmaryrd}
\usepackage{algorithmicx}
\usepackage[noend]{algpseudocode}

\usepackage{array}

\usepackage{mdwmath}
\usepackage{mdwtab}

\usepackage{caption}
\usepackage{subcaption}
\usepackage{stfloats}

\usepackage{url}

\newcommand{\AssociationStrength}[1]{\sigma_{#1}}
\newcommand{\paren}[1]{\left( #1 \right)}
\newcommand{\cut}[1]{}

\newcommand{\Aggressive}{24hr$\times$2}
\newcommand{\VeryConservative}{12hr$\times$1.5}
\newcommand{\HeavyStart}{24hr$\times$2+2start}
\newcommand{\HeavierStart}{30min$\times$2}

\newtheorem{definition}{Definition}

\begin{document}
%
\title{Spaced Repetition and Mnemonics Enable Recall of Multiple Strong Passwords}

\author{\IEEEauthorblockN{Jeremiah Blocki}
\IEEEauthorblockA{Computer Science Department\\
Carnegie Mellon University\\
Pittsburgh, PA 15213\\
Email: jblocki@cs.cmu.edu}
\and
\IEEEauthorblockN{Saranga Komanduri}
\IEEEauthorblockA{Computer Science Department\\
Carnegie Mellon University\\
Pittsburgh, PA 15213\\
Email: sarangak@cs.cmu.edu}
\and
\IEEEauthorblockN{Lorrie Cranor}
\IEEEauthorblockA{Computer Science Department\\
Carnegie Mellon University\\
Pittsburgh, PA 15213\\
Email: lorrie@cs.cmu.edu}
\and
\IEEEauthorblockN{Anupam Datta}
\IEEEauthorblockA{Computer Science Department\\
Carnegie Mellon University\\
Pittsburgh, PA 15213\\
Email: danupam@cs.cmu.edu}}

\IEEEoverridecommandlockouts
\makeatletter\def\@IEEEpubidpullup{9\baselineskip}\makeatother
\IEEEpubid{\parbox{\columnwidth}{Permission to freely reproduce all or part
    of this paper for noncommercial purposes is granted provided that
    copies bear this notice and the full citation on the first
    page. Reproduction for commercial purposes is strictly prohibited
    without the prior written consent of the Internet Society, the
    first-named author (for reproduction of an entire paper only), and
    the author's employer if the paper was prepared within the scope
    of employment.  \\
    NDSS '15, 8-11 February 2015, San Diego, CA, USA\\
    Copyright 2015 Internet Society, ISBN 1-891562-38-X\\
    http://dx.doi.org/10.14722/ndss.2015.23094
}
\hspace{\columnsep}\makebox[\columnwidth]{}}

\maketitle

\begin{abstract}
    We report on a user study that provides evidence that spaced repetition and a specific mnemonic technique
    enable users to successfully recall multiple strong passwords over time. Remote research participants were
    asked to memorize $4$ Person-Action-Object (PAO) stories where they chose a famous person from a drop-down
    list and were given machine-generated random action-object pairs. Users were
    also shown a photo of a scene and asked to imagine the PAO story taking place in the scene (e.g.,
    Bill Gates---swallowing---bike on a beach). Subsequently, they were asked to recall the action-object
    pairs when prompted with the associated scene-person pairs following a spaced repetition schedule over
    a period of $127+$ days. While we evaluated several spaced repetition schedules, the best results
    were obtained when users initially returned after $12$ hours and then in $1.5\times$ increasing intervals:
    $77\%$ of the participants successfully recalled all $4$ stories in $10$ tests over a period of $\approx 158$ days. Much of the forgetting happened in the first test period ($12$ hours):  $89\%$ of
    participants who remembered their stories during the first test period successfully remembered them in every subsequent round.
    These findings, coupled with recent results on naturally rehearsing password schemes, suggest that $4$ PAO stories
    could be used to create usable and strong passwords for $14$ sensitive
    accounts following this spaced repetition schedule, possibly with a few extra upfront rehearsals.
    In addition, we find statistically significant evidence that with $8$ tests over $64$ days
    users who were asked to memorize $4$ PAO stories outperform users who are given $4$ random action-object pairs,
    but with $9$ tests over $128$ days the advantage is not significant.
    Furthermore, there is an interference effect across multiple PAO stories: the recall rate of
$100\%$ (resp. $90\%$) for participants who were asked to memorize $1$ PAO story (resp. $2$ PAO stories) is significantly better than the rate for
participants who were asked to memorize $4$ PAO stories. These findings yield concrete advice for improving constructions of password management schemes
    and future user studies.


\end{abstract}


%

\section{Introduction}
Passwords are currently the dominant form of human authentication over the Internet despite many attempts to replace them \cite{bonneau2012quest}. A typical internet user has the complex task of creating and remembering passwords for many different accounts. Users struggle with this task, adopting insecure password practices \cite{florencio2007large,center2010consumer,kruger2008empirical,bonneau2012science} or frequently having to reset their passwords.
Yet research on human memory provides reason for optimism. Specifically, \emph{spaced repetition}---a memorization technique that
incorporates increasing intervals of time between subsequent review of previously learned material---has been
shown to be effective in enabling recall in a wide variety of
domains~\cite{memory:ebbinghaus1913,superMemo,wozniak2007supermemo,Pimsleur1967,memory:textbook:baddeley1997}.
Similarly, \emph{mnemonic} techniques that provide multiple semantic encodings of information (e.g., as stories and images)
also significantly help humans recall information~\cite{memory:textbook:baddeley1997,Memory:10000Pictures:standingt1973}.

We report on a user study that provides evidence that spaced repetition and mnemonics
enable users to successfully recall \emph{multiple} strong passwords over time. The study is inspired
by a recent result on \emph{naturally rehearshing password schemes}~\cite{blockiNaturallyRehearsingPasswords} that rely on spaced repetition
and a specific Person-Action-Object (PAO) mnemonic technique to design a scheme to create and maintain multiple strong passwords.
As a core component of the study, remote research participants were asked to memorize $4$ Person-Action-Object (PAO)
stories where they chose a famous person from a drop-down list and were given machine-generated random action-object pairs. Users were
also shown a photo of a scene and asked to imagine the PAO story taking place in the scene (e.g.,
Bill Gates---swallowing---bike on a beach). Subsequently, they were asked to recall the action-object
pairs (e.g., swallowing---bike) when prompted with the associated scene-person pairs (e.g., Bill Gates---beach)
following a spaced repetition schedule over a period of $100+$ days. We designed the study to seek answers to the following questions:
\begin{itemize}
    \item Do users who follow spaced repetition schedules successfully recall multiple PAO stories and, if so, which schedules
        work best?
    \item Does the PAO mnemonic technique improve recall over random action-object pairs alone?
    \item Is there an interference effect when users are asked to memorize multiple PAO stories?
\end{itemize}

We summarize our key findings and discuss their implications for password management below.
First, while we evaluated several spaced repetition schedules, the best results
were obtained under the schedule in which  users initially returned after $12$ hours and then in
$1.5\times$ increasing intervals: $76.6\%$ of the participants successfully recalled all $4$ stories in $10$ tests over a period of $\approx 158$ days. Much of the forgetting happened in the first test period (the first $12$ hours): $89\%$ of participants who remembered their stories during the first test period successfully remembered them in every subsequent round. This finding, coupled with recent results of Blocki et al.~\cite{blockiNaturallyRehearsingPasswords}, suggest that
$4$ PAO stories could be used to create and maintain usable and strong passwords for up to $14$
accounts following this spaced repetition schedule, possibly with a few extra upfront rehearsals.
The finding that much of the forgetting happens in the first test period robustly held in all the
spaced repetition schedules that we experimented with. Another implication of this finding is that
password expiration policies~\cite{guideline2006nist} negatively impact usability by
forcing users to return to the highest rehearsal effort region of memorizing a password.
Furthermore, they are unnecessary for strong passwords (see Section~\ref{sec:Background}).

Second, we find statistically significant evidence that initially with $8$ tests over $64$ days users
who were asked to memorize $4$ PAO stories outperform users who are given $4$ random action-object pairs,
but with $9$ tests over $128$ days the advantage is not significant. This finding is consistent with the
previous finding in that much of the forgetting happens in the early rounds and in those rounds the
PAO mnemonic technique helps significantly with recall.

Third, we find a statistically significant interference effect across multiple PAO stories. Specifically, the recall rate of
$100\%$ (resp. $90\%$) for participants who were asked to memorize $1$ PAO story (resp. $2$ PAO stories) is significantly better than the rate for
participants who were asked to memorize $4$ PAO stories.  The interference effect is strong: it continues to be statistically
significant even if we only count a participant with $4$ PAO stories as failing if they forgot their first (or first two)
action-object pair(s). This finding has several implications for password management. Further studies are needed to
discover whether the interference effect is alleviated if users memorize multiple PAO stories following a staggered
schedule in which they memorize $2$ stories at a time. To accommodate this user model, we also need new constructions for
naturally rehearsing password schemes in which passwords can be constructed even when not all PAO stories are memorized
upfront (see Section~\ref{sec:UserStudyDiscussion} for a concrete open problem).  At the same time, the perfect recall
rate for $1$ or $2$ PAO stories suggests that they could serve as a mechanism for strengthening existing passwords over time.
This conclusion is similar to the conclusion of a related study of Bonneau and Schechter~\cite{BS14} (although there are significant
differences between the two studies that we discuss in Section~\ref{sec:UserStudyRelated}).

\noindent
\emph{Organization.} Section \ref{sec:Background} briefly reviews the password management scheme of Blocki et al.
\cite{blockiNaturallyRehearsingPasswords}, and the security of the associated passwords consisting of random action-object pairs.
Section~\ref{sec:StudyDesign} presents the design of our user study. Section~\ref{sec:Results} describes the results of the study.
Section~\ref{sec:UserStudyRelated} describes related work.  Finally, Section \ref{sec:UserStudyDiscussion} concludes with a
discussion of the implications of these results for password management and suggestions for future work.

\section{Background} \label{sec:Background}
In this section we show how a user can form multiple secure passwords from a few random PAO stories by following the Shared Cues password management scheme of Blocki et al.\cite{blockiNaturallyRehearsingPasswords}. We first analyze the security of passwords consisting of randomly selected actions and objects in Section \ref{subsec:SecurityAgainstOfflineAttacks}. In Section \ref{subsec:SharedCues} we overview the Shared Cues password management scheme. In Section \ref{subsec:variant} we consider a variation of the Shared Cues password management scheme which only requires the user to memorize four PAO stories to form $14$ strong passwords.

\subsection{Security Against Offline Attacks} \label{subsec:SecurityAgainstOfflineAttacks} Any adversary who has obtained the cryptographic hash $\mathbb{H}\paren{pw}$ of a user's password $pw$ can mount an automated brute-force attack to crack the password by comparing $\mathbb{H}\paren{pw}$ with the cryptographic hashes of likely password guesses. This attack is called an offline dictionary attack, and there are many password crackers that an adversary could use \cite{JTR}. Offline dictionary attacks against passwords are powerful and commonplace \cite{PasswordCrackingArticle}.  Adversaries have been able to compromise servers at large companies (e.g., Zappos, LinkedIn, Sony, Gawker
\cite{breach:Zappos,breach:sony,breach:militaryHACK,breach:linkedin,breach:rockyou,breach:IEEE}) resulting in the release of millions of cryptographic password hashes.

In expectation an adversary would need to pay $N\mathbb{C}\paren{\mathbb{H}}/2$ to crack a password chosen uniformly at random from a space of size $N$, where  $\mathbb{C}\paren{\mathbb{H}}$ denotes the cost of evaluating the cryptographic hash function $\mathbb{H}$ one time. In our study each action is chosen uniformly at random from a list of $92$ actions and each object is chosen from a list of $96$ objects. Thus, $N = \paren{8,740}^i$ for $i$ randomly chosen action-object pairs (approximately equivalent to a randomly chosen $4$-digit pin number when $i=1$). Table \ref{tab:OfflineCost} shows the expected cost ($\paren{8,740}^i\mathbb{C}\paren{\mathbb{H}}/2$) of an offline attack against a password consisting of $i$ randomly chosen action-object pairs. Symantec reported that compromised passwords are sold for between \$4 and \$30 on the black market\cite{passwordBlackMarket}. As long as $\mathbb{C}\paren{\mathbb{H}} \geq \$10^{-6}$ a rational adversary would not bother trying to crack a password consisting of two random action-object pairs. A password consisting of three random action-object pairs would be strong enough to protect higher value accounts.

Bonneau and Schechter used data on the Bitcoin mining economy to estimate that $\mathbb{C}\paren{\mathbb{H}}  \approx \$ 1.2 \times 10^{-15}$ for the SHA-256 hash function, and they estimate that iterated password hashing\footnote{Cryptographic password hash functions like SCRYPT or BCRYPT\cite{bcrypt} use similar ideas to increase $\mathbb{C}\paren{\mathbb{H}}$ .} can only  increased this cost to $\mathbb{C}\paren{\mathbb{H}} \approx \$1.42 \times 10^{-8}$ --- unless we are willing to wait more than two seconds to compute $\mathbb{H}$ during authentication. However, we note that the value of $\mathbb{C}\paren{\mathbb{H}}$ could be increased without increasing authentication time using other techniques (parallel computation, memory hard functions). For example, if authentication were performed on a GPU with $1,024$ cores and we were willing to wait approximately two seconds for authentication we could increase $\mathbb{C}\paren{\mathbb{H}} \approx \$2^{-16.07}\approx \$1.46 \times 10^{-5}$ by developing a function $\mathbb{H}$ whose computation can be divided easily\footnote{One concrete way to accomplish this would be to set $\mathbb{H}(x) = \mathbb{H}_1\paren{ \varoplus_{y \in \{0,1\}^d} \mathbb{H}_2\paren{x,y} }$, where $\mathbb{H}_1$ and $\mathbb{H}_2$ are also cryptographic hash functions. Each of the $2^d$ calls to $\mathbb{H}_2$ could be evaluated in parallel.}.  

\subsection{Shared Cues Password Management Scheme} \label{subsec:SharedCues}
Our user study is partially motivated by the Shared Cues password management scheme of  Blocki et al.\cite{blockiNaturallyRehearsingPasswords}. In their scheme the user memorizes random PAO stories, and forms his passwords by appending the secret action(s) and object(s) from different stories together.

\noindent
\emph{Person-Action-Object Stories.} A user who adopts the Shared Cues password management scheme\cite{blockiNaturallyRehearsingPasswords} first memorizes several randomly generated Person-Action-Object (PAO) stories. To memorize each PAO story the user would be shown four images: a person, an action, an object and a scene. The user is instructed to imagine the PAO story taking place inside the scene. After the user has memorized a PAO story the computer stores the images of the person and the scene, but discards the images of the secret action and object. 

A password is formed by concatenating the secret action(s) and object(s) from several different PAO stories. During authentication the images of the corresponding people/scenes are used as a public cue to help the user remember his secret stories. The public cues remind the user which secrets are used to form each password (e.g., take the action from the PAO story involving Bill Gates on the beach, append the object from the PAO story involving Steve Jobs in the woods,...). The rehearsal phase from our user study emulates this authentication process. 

We stress that the actions and the objects in each of these PAO stories are selected uniformly at random by the computer after the images of the person/scene have been fixed. If the user selected the action and the object then he might pick actions or objects that are correlated with the person or the scene (e.g., users might favor objects like `apple' for a person like Steve Jobs). By having the computer select the story we ensure that the secret actions and objects are not correlated with the public cue for the password. Thus, an adversary who is able to observe these public cues does not gain any advantage in guessing the corresponding password. 

\noindent
\emph{Sharing Stories.} Stories are shared across different accounts to minimize the total number of stories that the user needs to remember and, more importantly, to maximize the natural rehearsal rate for each of the user's PAO stories. Blocki et al.\cite{blockiNaturallyRehearsingPasswords} proposed using a particular combinatorial design (definition \ref{def:GoodSharing}) to balance security and usability. This combinatorial design, which they called an $\left(n, \ell, \gamma \right)$-sharing set family, ensures that no pair of passwords can share too many of the same secret actions/objects. Thus, an adversary who has seen one of the user's passwords will not be able to guess {\em any} of the user's other passwords. More formally, let $S_i$ (resp. $S_j$) denote the subset of secrets (actions/objects) used to form the password $pw_i$ (resp. $pw_j$) for account $A_i$ (resp. $A_j$). Even if the adversary sees $pw_j$ he still has to guess all of the secrets in $S_i-\left(S_i \cap S_j\right)$ before he can obtain $pw_i$. In an $\left(n, \ell, \gamma \right)$-sharing set family we can ensure that the set $S_i-\left(S_i \cap S_j\right)$ contains at least $\ell-\gamma$ secrets.

\begin{definition} \label{def:GoodSharing} We say that a set family $\mathcal{S} = \{S_1,...,S_m\}$ is $\left(n, \ell, \gamma \right)$-sharing if (1) $\left| \bigcup_{i=1}^m S_i \right| = n$, (2)$\left| S_i \right| = \ell$ for each $S_i \in \mathcal{S}$, and (3) $\left| S_i \cap S_j \right| \leq \gamma$ for each pair $S_i \neq S_j \in \mathcal{S}$.   \end{definition}

Here, $n$ denotes the number of secrets (actions/objects) that the user has to memorize and $m$ denotes the number of passwords that the user can form. Intuitively, we want to keep $n$ as small as possible to minimize memorization effort. Even if the adversary learns the users password $pw_j$ for account $A_j$ then the password for account $S_i$ is still at least as strong as a password containing $\ell-\gamma$ secrets. 

Blocki et al.\cite{blockiNaturallyRehearsingPasswords} showed how a user could create $m=110$ unique passwords from $n=43$ PAO stories using a $(43,4,1)$-sharing set family (in their scheme each password consisted of four action-object pairs). Even if an adversary was able to obtain two of the user's 
passwords all of the user's remaining passwords would be at least as strong as a password containing $2$ random action-object pairs --- strong enough to resist offline attacks\footnote{\label{footnote:PwdCrack}Assuming that the passwords are encrypted with a cryptographic hash function $\mathbb{H}$ with $\mathbb{C}\paren{\mathbb{H}} \geq \$10^{-6}$ and that the adversary is not willing to spend more than $\$30$ cracking the password\cite{passwordBlackMarket}.}. Even if an adversary was able to obtain three of the user's passwords all of the user's remaining passwords would be strong enough to resist online attacks --- each remaining password contains at least one unknown action-object pair.

\emph{Usability Model.} Blocki et al.\cite{blockiNaturallyRehearsingPasswords} developed a usability model to predict how much work a user would need to do to remember all of his secrets. A central piece of their model was based on an assumption about human memory that they called the expanding rehearsal assumption. Loosely, this assumption states that a person will be able to remember his secrets if he follows a spaced repetition schedule like the ones tested in this study.

\subsection{Variants Considered in Our Study} \label{subsec:variant}
 We observe that $4$ PAO stories is already enough to generate $14$ moderately secure passwords in the Shared Cues framework using a $\paren{8,4,2}$-sharing set family of size $m=14$. To see this we observe that $\mathcal{S} = \{\{1, 2, 3, 4\}$, $\{1, 2, 5, 6\}$, $\{1, 2, 7, 8\}$, $\{1, 3, 5, 7\}$, $\{1, 3, 6,
  8\}$, $\{1, 4, 5, 8\}$, $\{1, 4, 6, 7\}$, $\{2, 3, 5, 8\},$ $\{2, 3, 6, 7\},$ $\{2, 4,
  5, 7\}$, $\{2, 4, 6, 8\}$, $\{3, 4, 5, 6\}$, $\{3, 4, 7, 8\}$, $\{5, 6, 7, 8\}\}$ is an $\paren{8,4,2}$-sharing set family. Each PAO story that the user memorizes can be viewed as containing two independent secrets (the action and the object). Thus, each password will contain $\ell=4$ secrets (actions and/or objects). The password $pw_1$ for the first account $A_1$ would be formed by appending the actions and objects from the first two PAO stories and the password $pw_{14}$ for account $A_{14}$ would be formed by appending the actions and objects from the last two PAO stories.

\noindent
\emph{Security.} Each password is strong enough to resist an offline attack (see \ref{footnote:PwdCrack}). Even if the adversary recovered one of the passwords in a plaintext password breach all of the user's other passwords are strong enough to resist online attacks\cut{\footnotemark[\ref{footnote:Online}]} because each password will contain at least two unknown secrets (action(s) and/or object(s))\cut{\footnote{We remark that there is a $\paren{8,4,3}$-sharing set family of size $m= 70 = {8 \choose 4}$ which would allow the user to generate $70$ passwords. However, in this case there is a good chance an adversary who has seen one of the user's plaintext passwords would be able crack at least one of the user's other passwords in an online attack --- even with a strict $3$ strikes policy at each account.}}.

\noindent
\emph{Usability.}
The evaluation of usability of this construction can be decomposed into two questions. First, can users robustly recall $4$ PAO
stories while following a suitable spaced repetition schedule? A central goal of our study is to answer this question. Second, how many
extra rehearsals (beyond rehearsals from normal logins) does a user have to perform in order to follow the spaced repetition schedule?
We do not attempt to answer this question in our study. However, we provide a sense of this user effort in the discussion section
(see Section~\ref{sec:UserStudyDiscussion}).

\begin{table}[t]
\centering
\begin{tabular}{|c|c|c|c|c|}
\hline
& \multicolumn{4}{c|}{\# Action-Object Pairs in Password} \\
\hline
$\mathbb{C}\paren{\mathbf{H}}$ & One & Two & Three & Four \\
\hline

$\$10^{-5}$ & $\$4.4\times 10^{-2}$ & $\$390$ & $\$3.4\times 10^{6}$ & $\$3.0 \times 10^{10}$ \\
\hline
$\$10^{-6}$ & $\$4.4\times 10^{-3}$ & $\$39$ & $\$3.4\times 10^{5}$ & $\$3.0 \times 10^{9}$ \\
\hline
$\$10^{-7}$ & $\$4.4\times 10^{-4}$ & $\$3.9$ & $\$3.4\times 10^{4}$ & $\$3.0 \times 10^{8}$ \\
\hline
\end{tabular}
\caption{Expected Cost of an Offline Attack}
\label{tab:OfflineCost}
\end{table}

\section{Study Design} \label{sec:StudyDesign}
Our user study was conducted online using Amazon's Mechanical Turk framework, on a website at our institution. It was approved by the Institutional Review Board (IRB) at Carnegie Mellon University. After participants consented to participate in the research study, we randomly assigned each participant to a particular study condition. Members in a particular condition were assigned a particular number of action-object pairs (either $1$, $2$, or $4$), a particular memorization technique (mnemonic or text), and a particular rehearsal schedule (e.g., \Aggressive, \VeryConservative) as determined by the condition.

Participants were then asked to complete a memorization phase. We randomly selected actions (e.g., swallowing) and objects (e.g., bike) for each participant to memorize. Participants in mnemonic conditions were also assigned pictures or ``scenes,'' one for each action-object pair, and were given specific instructions about how to memorize their words. We paid participants $\$0.50$ for completing the memorization phase. Once participants completed the memorization phase we asked them to return periodically to rehearse their words. To encourage participants to return we paid participants $\$0.75$ for each rehearsal, whether or not they were able to remember the words. If a participant forgot an action-object pair, then we reminded the participant of the actions and objects that were assigned and asked that participant to complete the memorization phase again. 

We restricted our participant pool to those Mechanical Turk workers who had an approval rate of 95\% or better, had completed at least 100 previous tasks, and were identified by Amazon as living in the United States. $797$ participants visited our study website, and $578$ completed the memorization phase and initial rehearsal phase.


\subsection{Recruitment} \label{subsec:RecruitmentText}
On the Mechanical Turk website, participants were recruited with the
following text:\\

\begin{quotation}
Participate in a Carnegie Mellon University research study on memory.
You will be asked to memorize and rehearse random words for a 50 cent
payment. After you complete the memorization phase, we will periodically
ask you to return to check if you still remember the words. If you forget the
words then we will remind you of the words and ask you to complete the
memorization phase again. You will be paid 75 cents upon the completion
of each rehearsal. 

Because this is a memory study we ask that you do not write down the
words that we ask you to memorize. You will be paid for each completed
rehearsal phase --- even if you forgot the words.
\end{quotation}


\subsection{Memorization Phase}

\subsubsection{Mnemonic group}
We first describe the memorization phase for participants assigned to a mnemonic condition. Participants in the mnemonic group were given the following instructions:
 
\begin{quotation}
This study is being conducted as part of a Carnegie Mellon University research project. It is important that you answer questions honestly and completely. Please take a minute to read the following instructions. 

The goal of this study is to quantify the effects of rehearsal and the use of mnemonic techniques on long term memory retention. In this study you will be asked to memorize and rehearse eight random words (four actions and four objects). During the first phase we will ask you to memorize the eight random words --- you will be paid $\$0.50$ upon completion of the memorization phase. After you complete the memorization phase we will periodically ask you to return via email to check if you still remember the words. If you forget the words, we will remind you of the words and ask you to complete the memorization phase again. You will be paid $\$0.75$ upon the completion of each rehearsal. 

{\bf Important:} Because this is a memory study we ask that you do not write down the words we ask you to memorize. You will be paid for each completed rehearsal phase --- even if you forgot the words. 

You have been assigned to the mnemonic group, which means that we give you specific instructions about how to memorize the words. One of the purposes of this study is to determine how effective certain mnemonic techniques are during the memorization task. We ask that you follow the directions exactly --- even if you would prefer to memorize the words in a different way. 
\end{quotation}

\begin{figure}[tb]
\centering
\includegraphics[scale=0.30]{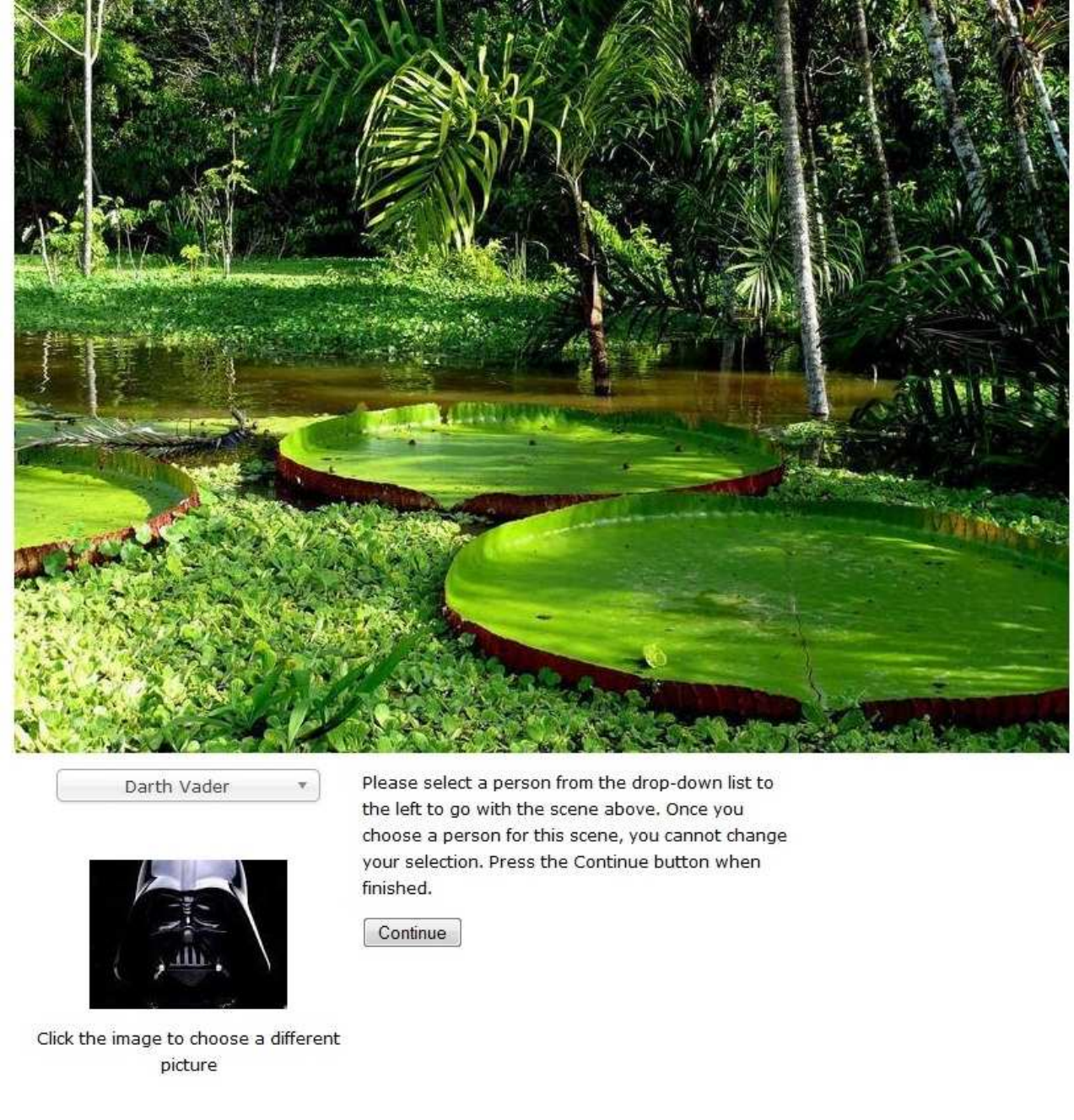}
\caption{Memorization Step 1. Scene and Person.}
\label{fig:BillGates}
\end{figure}

After participants finished reading the instructions the memorization phase proceeded as follows:

\noindent {\bf Step\:1} Initially, participants were shown a photo of an assigned scene. Participants were next asked to select a famous person or character from a predefined list (e.g., Darth Vader) and were shown a photograph of the famous person that they selected --- see Figure \ref{fig:BillGates}.  Once selected, participants could not change their person choices.  After selecting a person, participants saw a randomly selected action-object pair. 
See Appendix \ref{subsec:ListOfPAO} for the lists of people, actions and objects used in the study. 

\begin{figure}[tb]
\centering
\includegraphics[scale=0.3]{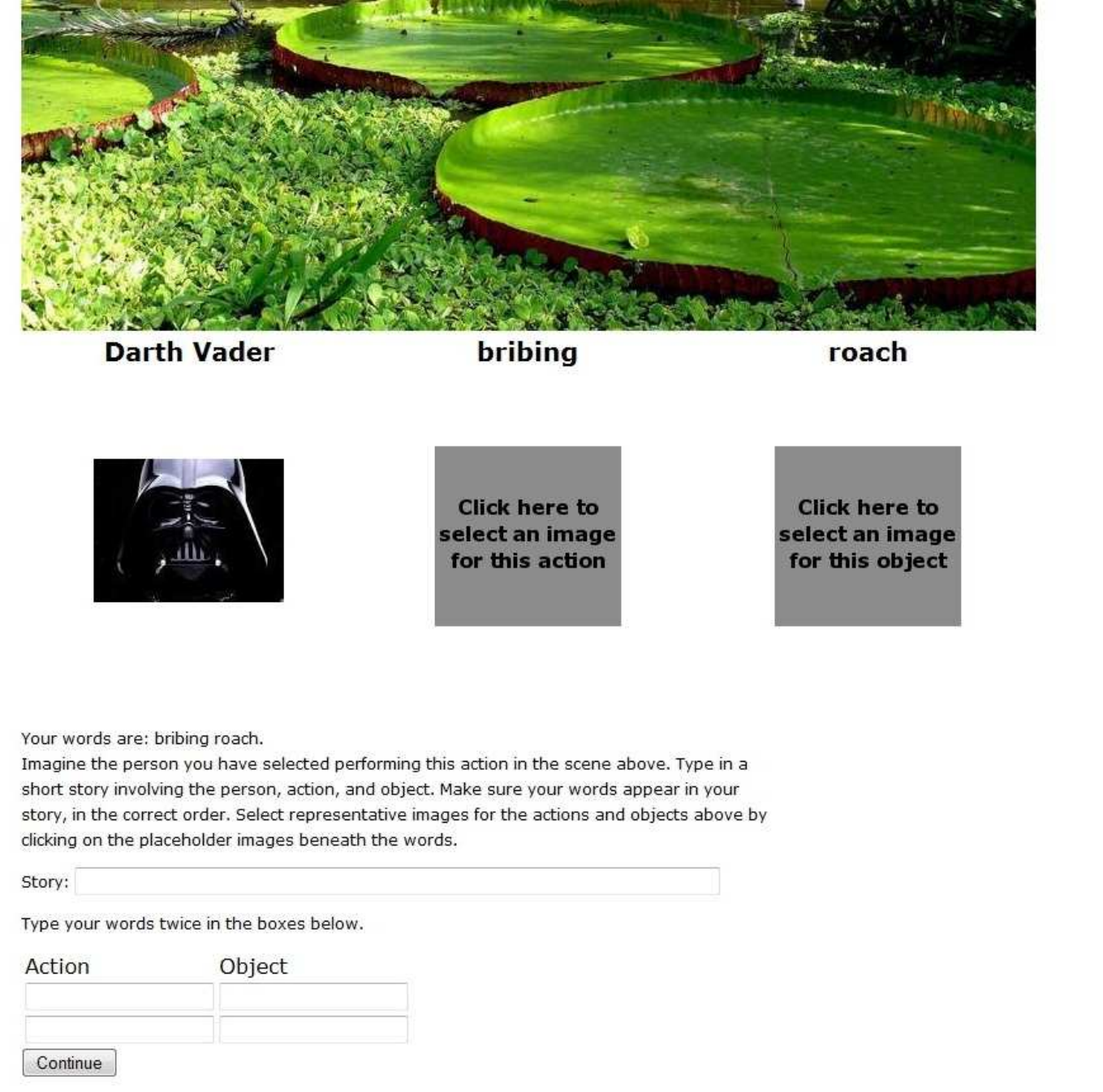}
\caption{Memorization Steps 2--3. Darth Vader bribing a roach on the lily pad.} \label{fig:UserStudyPAOStory}
\end{figure}

\noindent {\bf Step\:2} As shown in Figure \ref{fig:UserStudyPAOStory}, we asked participants to imagine a story in which the person they selected is performing the action in the given scene 
(e.g., imagine Darth Vader bribing the roach on the lily pad). We asked participants to type in this story, with all words in the correct order (Person-Action-Object). 

\noindent {\bf Step\:3} Participants were then required to select photographs of the action and object, and type in the action and object words two more times in separate fields.

We asked most participants to repeat Steps 1 through 3 four times using a new scene (e.g., a baseball field or a hotel room underneath the sea), a new famous person/character and a new action-object pair during each repetition. Thus, most participants memorized a total of eight words (four actions and four objects). After the memorization phase, we asked participants to complete a rehearsal phase (See Figure \ref{fig:rehearsal}) before leaving the website.

\subsubsection{Text group}
\begin{figure}[tb]
\centering
\includegraphics[scale=0.5]{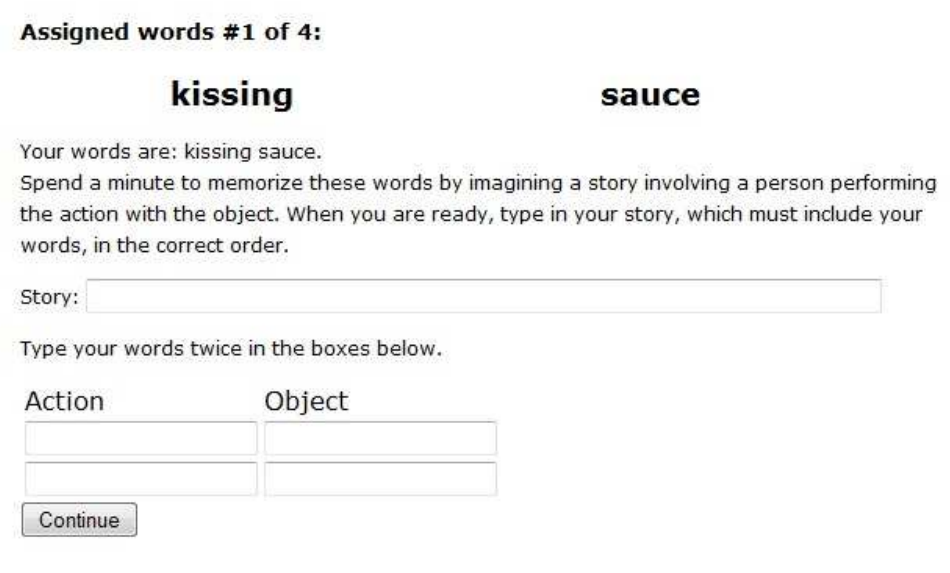}
\caption{Memorization Steps 1--3 for Text group.}
\label{fig:UserStudyNonMnemonicGroup}
\end{figure}

We next describe the memorization phase for participants assigned to a text condition. Participants in the text group were given the same instructions as those in the mnemonic group, with the exception of the last paragraph, which begins ``You have been assigned to the mnemonic group\dots''  This paragraph is omitted for those in a text condition. After participants finished reading the instructions, the memorization phase proceeded as follows:

\noindent{\bf Step\:1} As shown in Figure \ref{fig:UserStudyNonMnemonicGroup}, we randomly selected an action-object pair, and displayed these words to the participant.

\noindent{\bf Step\:2} We asked each participant to spend one minute memorizing his words. We suggest that participants imagine a person performing the action with the object.  We asked each participant to type in a story which includes the action and the object in the correct order. 

\noindent{\bf Step\:3} Participants were then required to type in the action and object words two more times in separate fields.

As with the mnemonic group, we asked most participants to repeat Steps 1 through 3 four times, and asked participants to complete a rehearsal phase.

\subsection{Rehearsal Phase}
Each participant was assigned a particular rehearsal schedule. The particular times that we ask the participant to return were given by the rehearsal schedule that participant was assigned to use (see Table \ref{tab:Rehearsal}). We e-mailed participants to remind them to return for each rehearsal using the following text:

\begin{quotation}
Dear Carnegie Mellon study participant:
Please return to (url) to participate in the next part of the memory study. If you do not return promptly upon receiving this email, you might not be considered for future phases of the study. You will receive a $\$0.75$ bonus payment for completing this task and it should take less than five minutes.

Remember that you should not write down the words that were assigned to you. You will be paid for each completed rehearsal phase --- even if you forgot the words.

...

If, for any reason, you do not want to complete the study, please reply to this email and let us know why, so we can improve our protocol for future studies.

Thank you!

The Carnegie Mellon University Study Team
\end{quotation}

We describe the rehearsal phase below: 

\begin{figure}[tb]
\centering
\includegraphics[scale=0.3]{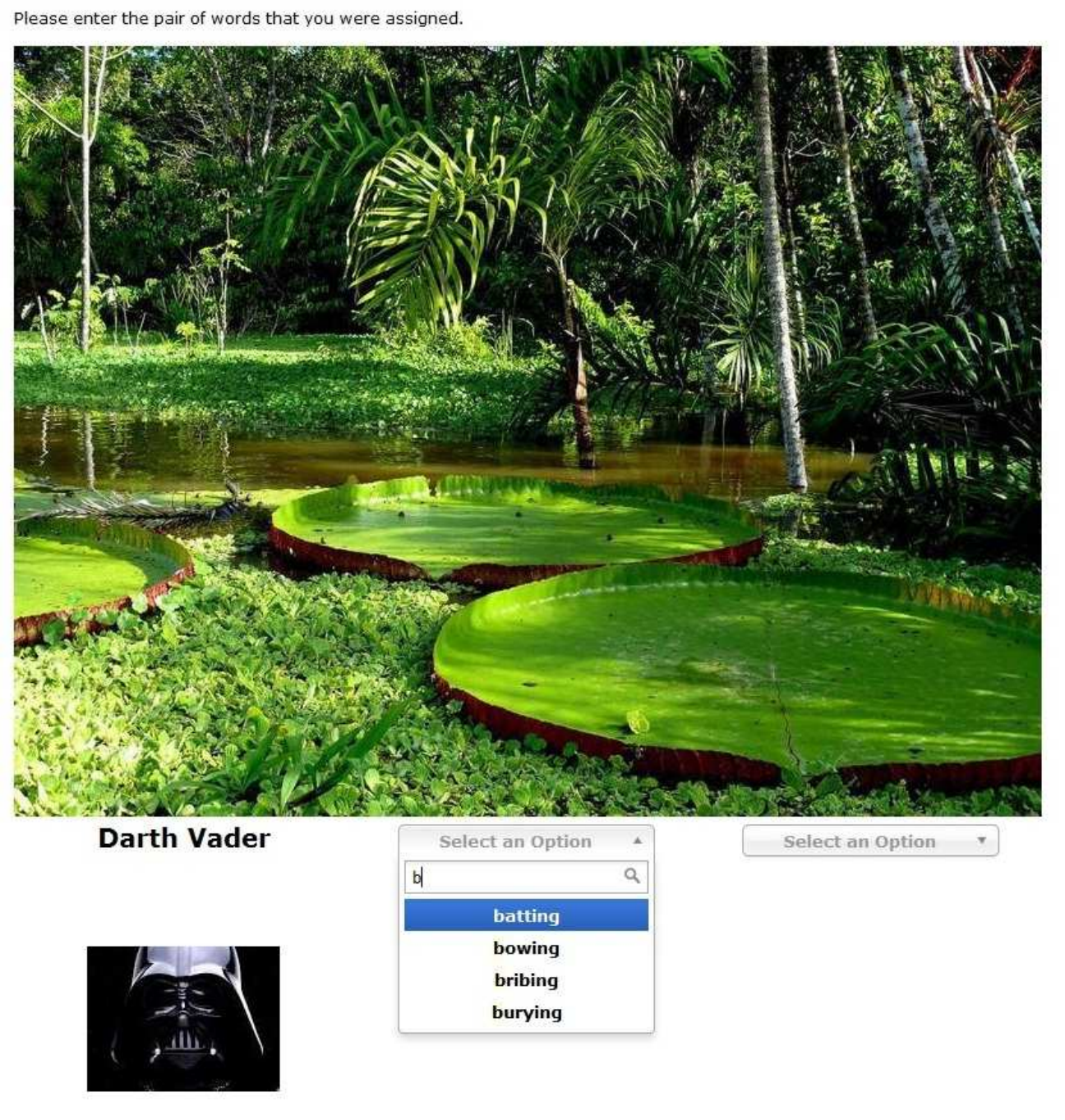}
\caption{Rehearsal Phase. Darth Vader and the photo of the lily pads on the Amazon River provide a cue to aid memory recall.} \label{fig:rehearsal}
\end{figure}

\subsubsection{Mnemonic group}
Each participant was shown the scene and the picture of the person that he chose while memorizing his first story during the memorization phase. We then asked each participant to recall the assigned action and object for that story. As shown in Figure \ref{fig:rehearsal}, actions and objects were browsable and searchable to aid recall.  If the participant was correct then we moved on to the next story. If the participant was incorrect then we asked the participant to try again. Each participant was allowed three guesses per action-object pair. After three incorrect guesses we asked the participant to repeat the memorization phase with the same actions and objects, and try the rehearsal phase again. Once the participant correctly entered all assigned action-object pairs, the rehearsal phase ended and participants were paid automatically.

\subsubsection{Text group} 
Each participant from the text group was simply asked to recall the actions and objects assigned during the memorization phase. As with the mnemonic group, actions and objects were browsable and searchable to aid recall.  Scoring and payment were handled the same as in the mnemonic conditions.

\subsection{Follow Up Survey}
Some participants did not return to rehearse their stories during the rehearsal phase. We cannot tell whether or not these participants would have remembered their passwords if they had returned. Instead we can only report the fraction of participants who remembered their passwords among those who returned for each rehearsal during the study. There are several reasons why a participant may not have returned (e.g., too busy, did not get the follow up message in time, convinced he or she would not remember the password). If participants do not return because they are convinced that they would not remember the password then this could be a source of bias (i.e., we would be selecting participants who are confident that they remember the story). Our hypothesis is that the primary reason that participants do not return is because they are too busy, because they did not get our follow up message in time, or because they are not interested in interacting with us outside of the initial Mechanical Turk task, and not because they were convinced that they would not remember the story. In order to test our hypothesis we sent a follow up survey to all participants who did not return to complete a rehearsal phase. 
Participants were paid 25 cents for completing this survey. The survey is described below:

\begin{quotation}
You are receiving this message because you recently participated in a CUPS Memory Study at CMU. A while ago you received an e-mail to participate in a follow up test. We would like to ask you to complete a quick survey to help us determine why participants were not able to return to complete this follow up study. The survey should take less than a minute to complete, and you will be paid 25 cents for completing the survey. The survey consists of one question. Which of the following reasons best describes why you were unable to return to take the follow up test?
\begin{enumerate}
\item[A] I no longer wished to participate in the study.
\item[B] I was too busy when I got the e-mail for the follow up test.
\item[C] I did not see the e-mail for the follow up test until it was too late.
\item[D] I was convinced that I would not be able to remember the words/stories that I memorized when I received the e-mail for the follow up test.
\item[E] I generally do not participate in follow up studies on mechanical turk.
\end{enumerate}
\end{quotation}

It is possible that some participants will choose not to participate in the follow up survey. However, in our case their decision not to participate is valuable information which supports our hypothesis, i.e., they are not interested in interacting with us outside of the initial Mechanical Turk task.

\subsection{Rehearsal Schedules} \label{sec:RehearsalSched}
In our study each participant was randomly assigned to follow one of the following rehearsal schedules \Aggressive, \HeavyStart, \HeavierStart{} and \VeryConservative. The specific rehearsal times for each schedule can be found in Table \ref{tab:Rehearsal}. We can interpret a schedule \Aggressive{} as follows: the length of the first rehearsal interval (e.g., the time between initial visit and the first rehearsal) is $24$ hours and the length of the $i+1$'th rehearsal interval is twice the length of the $i$'th rehearsal interval. If the participant was assigned to the \Aggressive{} rehearsal schedule then we would send that participant a reminder to rehearse $1$ day after the memorization phase and if that participant returned to complete the first rehearsal phase then we would send that participant another reminder to rehearse $2$ days after the first rehearsal. The next reminder would come four days later, etc. The final rehearsal would take place on day $1+2+...+32+64 = 127$. In the \VeryConservative{} schedule the length of the first rehearsal interval is $12$ hours and after that intervals grow by a factor of $1.5$. The \HeavyStart{} and \HeavierStart{} conditions are similar to the \Aggressive{} schedule except that participants are asked to do a few additional rehearsals on day 1 --- after this the rehearsal intervals are identical to the \Aggressive{} schedule. 

If a participant did not return to complete a particular rehearsal round then they were not asked to participant in subsequent rounds. We stress that this is the only reason why participants were dropped from the study. A participant who forgot one or more of his words during rehearsal $i$ would be reminded of his secret words and  would still be invited to participate in rehearsal $i+1$. 

We use the following syntactic pattern to denote a study condition: (Memorization Technique)\_(Rehearsal Schedule)\_(Number of action-object pairs memorized). For memorization technique we use $m$ to denote the mnemonic groups and $t$ to denote the text group. Thus, a participant in the group m\_\Aggressive\_4 refers to a user who was asked to memorize four actions and four objects using the mnemonic techniques we suggested and to rehearse his person-action-object stories following the \Aggressive{} rehearsal schedule from Table \ref{tab:Rehearsal}. Because most participants were asked to memorize four random action-object pairs we typically drop the ``\_4'' from the end of these conditions.

\begin{table*}[t]
\centering
\begin{tabular}{| p{1.in} | p{0.7in} | p{.4in} | p{2in} | p{2in} | }
\hline
{\bf Schedule} & {\bf Multiplier} & {\bf Base} & {\bf Rehearsal Intervals} & {\bf Rehearsal Days} \\
\hline 
\Aggressive & $\times 2$ & 1 Day &  1, 2, 4, 8, 16, 32, 64 & 1, 3, 7, 15, 31, 63, 127 \\
\hline 
\VeryConservative & $\times 1.5$ & 0.5 days & 0.5, 1.25, 2.4, 4, 6.5, 10,16,24,37,56 & 0.5, 1.75, 4.15, 8.15, 14.65, 24.65, 40.65, 64.65, 101.65, 157.65   \\
\hline
{ \HeavyStart }  & $\times 2$ & 1 Day  & 0.1 days, 0.5, 1, 2, 4, 8, 16, 32, 64 & 0.1, 0.6, 1.6, 3.6, 7.6, 15.6, 31.6, 63.6, 127.6  \\
\hline
{ \HeavierStart }  & $\times 2$ & 30 min  & 0.5 hr, 1 hr, 2 hr, 4 hr, 8 hr, 1 day, 2, 4, 8, 16, 32, 64  & 0.5hr, 1.5hr, 3.5hr, 7.5hr, 15.5hr, 1.65 days, 3.65, 7.65, 15.65, 31.65, 63.65, 127.65 \\
\hline
\end{tabular}
\caption{Rehearsal Schedules}
\label{tab:Rehearsal}
\end{table*}

\subsection{Online studies}
The passwords in our study did not protect high-value accounts,
limiting ecological validity. In contrast to real-world,
high-value passwords, study participants would not suffer
consequences beyond a modest time cost if they forgot their
password, nor were they incentivized to keep their passwords
only in memory beyond our repeated requests that they do so.

We recruited participants using Mechanical Turk (MTurk).
Using MTurk allows us to study a larger volume of participants in a
controlled setting
than would otherwise be possible.
MTurk workers tend to be younger, more educated, and more technical
than the general population,
but they represent a significantly more diverse population than is typically
used in lab studies, which often rely on college-student
participants~\cite{buhr+2011, Ipeirotis:2010}.
Many researchers have found that well-designed MTurk studies provide
high-quality user data~\cite{Downs:2010, Kittur:2008, Toomim:2011,
berinsky:2011, goodman:2012, horton:2010}.
Adar has criticized MTurk studies in general, although our use of
crowdsourcing to understand human behavior fits his description of an
appropriate use~\cite{adar:2011}.

\section{Results} \label{sec:Results}
In this section we present the results from our study. In Section \ref{subsec:studydata} we overview the raw data from our study (e.g., how many participants returned for each rehearsal round?) and some simple metrics (e.g., how many of these participants remembered their action-object pairs?)
In Section \ref{subsubsec:SurveyResults} we discuss the results of a survey we sent to participants that did not return for a rehearsal phase.
In Section \ref{sec:CoxRegression} we briefly overview Cox regression --- a tool for performing survival analysis that we used to compare several of our study conditions. In Section \ref{subsec:findings} we use the data from our study to evaluate and compare different study conditions.

\subsection{Study Data} \label{subsec:studydata}
\begin{table*}[htb]
\centering
 \begin{tabular}{|p{2cm}||p{0.58cm}|c|c|c|c|c|c|c|c|c|} 
\hline
                & {\bf Initial} &\multicolumn{9}{c|}{\begin{tabular}[x]{@{}c@{}}$\mathbf{NumSuccessfulReturned}\paren{i}$,\\$\mathbf{NumSurvived}(i)/\mathbf{NumSuccessfulReturned}\paren{i}$,\\  $95\%$ confidence interval\end{tabular}   } \\
\hline \begin{tabular}[x]{@{}c@{}}{\bf Rehearsal~$ i\backslash$}\\{\bf Condition}\end{tabular}
 &  $i$ $=$ $0$ & $1$ & $2$ & $3$ & $4$ & $5$ & $6$ & $7$ & $8$ & $9$ \\
\hline
m\_\HeavyStart  & 80 & \begin{tabular}[x]{@{}c@{}}51\\$94.1\%$ \\ 0.838,0.988 \end{tabular}  & \begin{tabular}[x]{@{}c@{}}41\\$100\%$ \\ 0.914,1 \end{tabular} & \begin{tabular}[x]{@{}c@{}}38\\$100\%$ \\ 0.907,1 \end{tabular} & \begin{tabular}[x]{@{}c@{}}37\\$97.3\%$ \\ 0.858,0.999 \end{tabular} & \begin{tabular}[x]{@{}c@{}}36\\$100\%$ \\ 0.903,1 \end{tabular} & \begin{tabular}[x]{@{}c@{}}36\\$97.2\%$ \\ 0.855,0.999 \end{tabular} & \begin{tabular}[x]{@{}c@{}}34\\$97.1\%$ \\ 0.847,0.999 \end{tabular} & \begin{tabular}[x]{@{}c@{}}30\\$90\%$ \\ 0.735,0.978 \end{tabular} & \begin{tabular}[x]{@{}c@{}}25\\$68\%$ \\ 0.465,0.85 \end{tabular} \\
\hline
t\_\HeavyStart
  & 100  &\begin{tabular}[x]{@{}c@{}}71\\$88.7\%$ \\ 0.790,0.950 \end{tabular}   &  \begin{tabular}[x]{@{}c@{}}54\\$96.3\%$ \\ 0.873,0.995 \end{tabular}  & \begin{tabular}[x]{@{}c@{}}51\\$100\%$ \\ 0.930,1 \end{tabular} & \begin{tabular}[x]{@{}c@{}}51\\$100\%$ \\ 0.930,1 \end{tabular} & \begin{tabular}[x]{@{}c@{}}51\\$94.1\%$ \\ 0.836,0.988 \end{tabular} & \begin{tabular}[x]{@{}c@{}}48\\$95.8\%$ \\ 0.857,0.995 \end{tabular} & \begin{tabular}[x]{@{}c@{}}44\\$88.6\%$ \\ 0.754,0.962 \end{tabular} & \begin{tabular}[x]{@{}c@{}}39\\$79.5\%$ \\ 0.635,0.907 \end{tabular} & 
\begin{tabular}[x]{@{}c@{}}39\\$86.2\%$ \\ 0.683,0.961 \end{tabular}   \\
\hline
m\_\Aggressive
 & 75  & \begin{tabular}[x]{@{}c@{}}65\\$76.9\%$ \\ 0.648,0.845 \end{tabular}  & \begin{tabular}[x]{@{}c@{}}45\\$93.3\%$ \\ 0.817,0.986 \end{tabular} & \begin{tabular}[x]{@{}c@{}}40\\$100\%$ \\ 0.912,1 \end{tabular}  & \begin{tabular}[x]{@{}c@{}}39\\$97.4\%$ \\ 0.865,0.999 \end{tabular}  & \begin{tabular}[x]{@{}c@{}}37\\$97.3\%$ \\ 0.858,0.999 \end{tabular}   & \begin{tabular}[x]{@{}c@{}}32\\$96.9\%$ \\ 0.838,0.999 \end{tabular}  &
\begin{tabular}[x]{@{}c@{}}28\\$89.3\%$ \\ 0.718,0.977 \end{tabular}  & N/A & N/A  \\
\hline
m\_\HeavyStart\_2
  & 81 &  \begin{tabular}[x]{@{}c@{}}50\\$100\%$ \\ 0.929,1 \end{tabular}  & \begin{tabular}[x]{@{}c@{}}42\\$100\%$ \\ 0.916,1 \end{tabular}    & \begin{tabular}[x]{@{}c@{}}42\\$100\%$ \\ 0.916,1 \end{tabular} & \begin{tabular}[x]{@{}c@{}}41\\$100\%$ \\ 0.914,1 \end{tabular}  &\begin{tabular}[x]{@{}c@{}}38\\$100\%$ \\ 0.907,1 \end{tabular} & \begin{tabular}[x]{@{}c@{}}37\\$100\%$ \\ 0.905,1 \end{tabular}   & \begin{tabular}[x]{@{}c@{}}36\\$100\%$ \\ 0.903,1 \end{tabular}  & \begin{tabular}[x]{@{}c@{}}33\\$100\%$ \\ 0.894,1 \end{tabular} & \begin{tabular}[x]{@{}c@{}}30\\$90\%$ \\ 0.735,0.979 \end{tabular} \\
\hline
m\_\HeavyStart\_1
  & 86  & \begin{tabular}[x]{@{}c@{}}64\\$100\%$ \\ 0.943,1 \end{tabular}     & \begin{tabular}[x]{@{}c@{}}52\\$100\%$ \\ 0.932,1 \end{tabular}   & \begin{tabular}[x]{@{}c@{}}49\\$100\%$ \\ 0.927,1 \end{tabular}    & \begin{tabular}[x]{@{}c@{}}49\\$100\%$ \\ 0.927,1 \end{tabular}  & \begin{tabular}[x]{@{}c@{}}47\\$100\%$ \\ 0.925,1 \end{tabular}   &  \begin{tabular}[x]{@{}c@{}}46\\$100\%$ \\ 0.923,1 \end{tabular}    & \begin{tabular}[x]{@{}c@{}}45\\$100\%$ \\ 0.922,1 \end{tabular}   & \begin{tabular}[x]{@{}c@{}}44\\$100\%$ \\ 0.920,1 \end{tabular}  & 
\begin{tabular}[x]{@{}c@{}}43\\$100\%$ \\ 0.918,1 \end{tabular}  \\
\hline
m\_\VeryConservative  & 83  & \begin{tabular}[x]{@{}c@{}}72\\$86.1\%$ \\ 0.759,0.931 \end{tabular}   & \begin{tabular}[x]{@{}c@{}}53\\$98.1\%$ \\ 0.899,1.000 \end{tabular}  &  \begin{tabular}[x]{@{}c@{}}51\\$100\%$ \\ 0.930,1 \end{tabular}  & \begin{tabular}[x]{@{}c@{}}51\\$100\%$ \\ 0.930,1 \end{tabular}  & \begin{tabular}[x]{@{}c@{}}49\\$100\%$ \\ 0.927,1 \end{tabular} & \begin{tabular}[x]{@{}c@{}}46\\$97.8\%$ \\ 0.885,0.999 \end{tabular}  & \begin{tabular}[x]{@{}c@{}}43\\$100\%$ \\ 0.918,1 \end{tabular}   & \begin{tabular}[x]{@{}c@{}}42\\$97.6\%$ \\ 0.874,0.999 \end{tabular} &  \begin{tabular}[x]{@{}c@{}}42\\$94.9\%$ \\ 0.827,0.994 \end{tabular}  \\
\hline
m\_\HeavierStart  & 73 & \begin{tabular}[x]{@{}c@{}}40\\$95\%$ \\ 0.831,0.994 \end{tabular}  &  \begin{tabular}[x]{@{}c@{}}27\\$100\%$ \\ 0.872,1 \end{tabular}  & \begin{tabular}[x]{@{}c@{}}26\\$100\%$ \\ 0.868,1 \end{tabular}  & \begin{tabular}[x]{@{}c@{}}24\\$100\%$ \\ 0.858,1 \end{tabular}  & \begin{tabular}[x]{@{}c@{}}22\\$100\%$ \\ 0.846,1 \end{tabular}  & \begin{tabular}[x]{@{}c@{}}22\\$100\%$ \\ 0.846,1 \end{tabular} & \begin{tabular}[x]{@{}c@{}}22\\$100\%$ \\ 0.846,1 \end{tabular} & \begin{tabular}[x]{@{}c@{}}22\\$100\%$ \\ 0.846,1 \end{tabular} & \begin{tabular}[x]{@{}c@{}}22\\$100\%$ \\ 0.846,1 \end{tabular} \\
\hline
\hline \begin{tabular}[x]{@{}c@{}}{\bf Rehearsal~$ i\backslash$}\\{\bf Condition}\end{tabular}
 &  &  $i$ $=$ $10$ &  $i$ $=$ $11$ & $i=12$ &  &  & &  &  &  \\
\hline
m\_\HeavierStart & & \begin{tabular}[x]{@{}c@{}}21\\$95.2\%$ \\ 0.762,0.999 \end{tabular} & \begin{tabular}[x]{@{}c@{}}20\\$90\%$ \\ 0.683,0.988 \end{tabular} & \begin{tabular}[x]{@{}c@{}}17\\$93.3\%$ \\ 0.681,0.998 \end{tabular} & & & & & & \\
 \hline
m\_\VeryConservative &  & \begin{tabular}[x]{@{}c@{}}36\\$100\%$ \\ 0.903,1 \end{tabular} & N/A  & N/A &  &  & &  &  &  \\
\hline 
\end{tabular}
\caption{$\mathbf{NumSurvived}(i)$/$\mathbf{NumSuccessfulReturned}(i)$ with $95\%$ binomial confidence intervals. $m=$ ``mnemonic," $t = $``text"}
\label{tab:remembered}
\end{table*}

One of the primary challenges in analyzing the results from our study is that some participants were dropped from the study because they were unable to return for one of their rehearsals in a timely manner. We do not know how many of  these participants would have been able to remember their stories under ideal circumstances. We consider several different ways to estimate the true survival rate in each condition. Before we present these metrics we must introduce some notation.

{\noindent \bf Notation:} We use $\mathbf{NumRemembered}\paren{C,i}$ to denote the number of participants from study condition $C$ who remembered their secret action-object pair(s) during rehearsal $i$  with $< 3$ incorrect guesses per action-object pair, and we use $\mathbf{NumSurvived}\paren{C, i}$ to denote the number of participants who also remembered their action-object pair(s) during every prior rehearsal. We use $\mathbf{NumReturned}\paren{C,i}$ to denote the total number of participants who returned for rehearsal $i$ and we use $\mathbf{NumSuccessfulReturned}\paren{C,i}$ to denote the total number of participants who survived through rehearsal $i-1$ and returned for rehearsal $i$. Finally, we use $\mathbf{Time}\paren{C,i}$ to denote the time of rehearsal $i$, as measured from the initial memorization phase. Because the study condition $C$ is often clear from the context we will typically omit it in our presentation.

\begin{figure}[tb]
\centering
\begin{subfigure}[b]{0.45 \textwidth}
\includegraphics[scale=0.4]{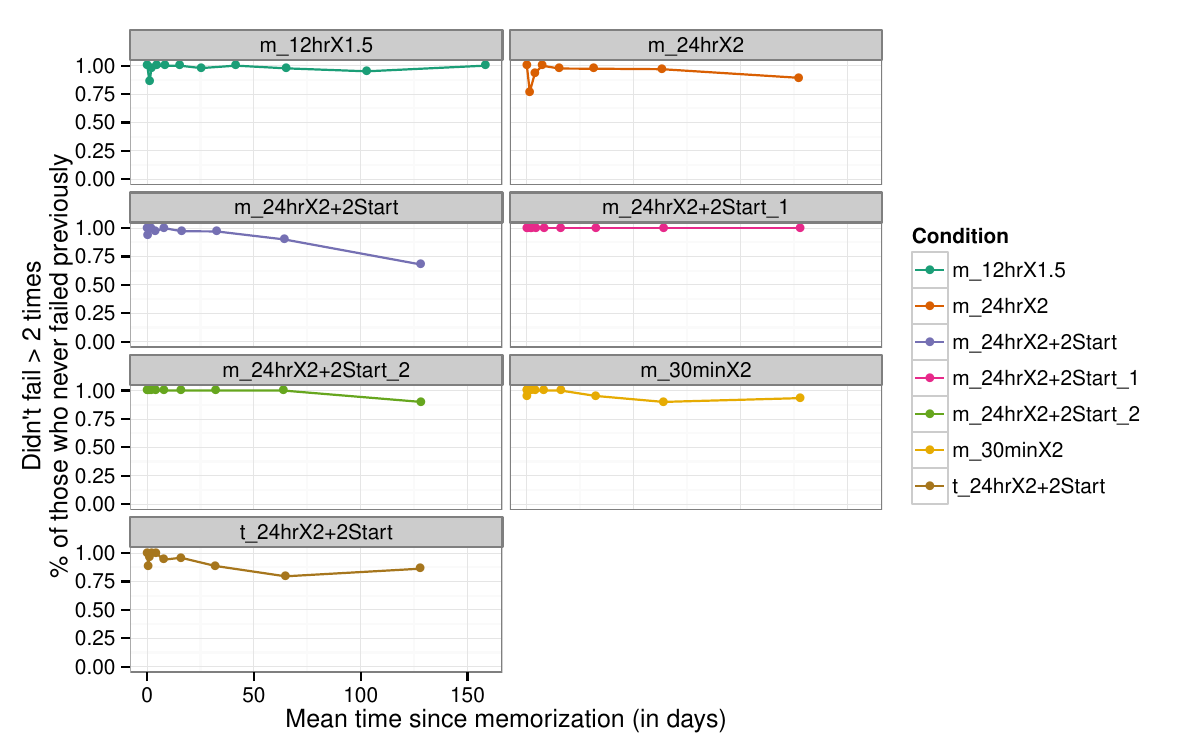}
\caption{Faceted. Mean Time Since Memorization.}
\label{fig:FacetedConditionalSurvivalProbability}
\end{subfigure}
\begin{subfigure}[b]{0.45 \textwidth}
\centering
\includegraphics[scale=0.4]{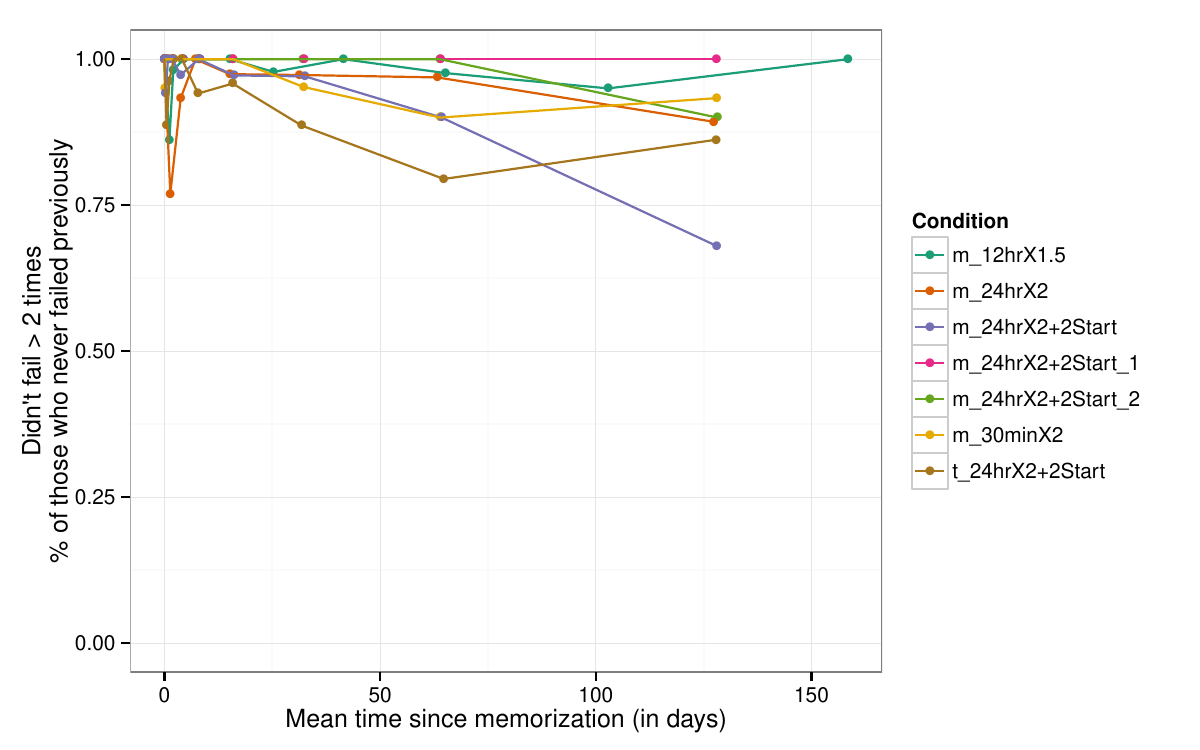}

\caption{Together. Mean Time Since Memorization. }
\label{fig:ConditionalSurvivalProbability}
\end{subfigure}
\caption{$\mathbf{NumSurvived}\paren{i}/\mathbf{NumSuccessfulReturned}\paren{i}$ vs.  $\mathbf{Time}\paren{i}$}
\end{figure}

Observe that $\frac{\mathbf{NumSurvived}\paren{i}}{\mathbf{NumSuccessfulReturned}\paren{i}}$ denotes the conditional probability that a participant remembers his PAO stories during rehearsal $i$ given that s/he has survived through rehearsal $i-1$ and returned for rehearsal $i$. Figures \ref{fig:FacetedConditionalSurvivalProbability} and  \ref{fig:ConditionalSurvivalProbability} plot the conditional probability of survival for participants in different study conditions. Table \ref{tab:remembered} shows how many participants who had never failed before returned in each rehearsal round as well as their conditional probability of success with  $95\%$ confidence intervals.

\begin{table}[t]
\centering
\resizebox{\columnwidth}{!}{%
 \begin{tabular}{|p{1.9cm}||p{0.5cm}|p{1.05cm}|p{1.05cm}|p{1.05cm}|p{1.05cm}|} 
\hline
                & {\bf Initial} &\multicolumn{4}{c|}{\begin{tabular}[x]{@{}c@{}}$\mathbf{Returned}\paren{i}$,\\$\mathbf{NumSurvived}(i)/\mathbf{NumReturned}(i)$,\\  $95\%$ confidence interval\end{tabular}   } \\
\hline \begin{tabular}[x]{@{}c@{}}{\bf Rehearsal~$ i\backslash$}\\{\bf Condition}\end{tabular}
 &  i$=$0 &\centering  \cut{$1$ & $2$ & $3$ & $4$ & $5$ & $6$ &\centering} $7$ &\centering  $8$ &\centering $9$ & \centering \arraybackslash $10$ \\
\hline
m\_\HeavyStart  & 80 & \cut{\begin{tabular}[x]{@{}c@{}}51\\$94.1\%$ \\ 0.838,0.988 \end{tabular}  & \begin{tabular}[x]{@{}c@{}}44\\$93.2\%$ \\ 0.813,0.986 \end{tabular} & \begin{tabular}[x]{@{}c@{}}41\\$92.7\%$ \\ 0.801,0.985 \end{tabular} & \begin{tabular}[x]{@{}c@{}}40\\$90\%$ \\ 0.763,0.97.2 \end{tabular} & \begin{tabular}[x]{@{}c@{}}40\\$90\%$ \\  0.763,0.97.2  \end{tabular} & \begin{tabular}[x]{@{}c@{}}40\\$87.5\%$ \\ 0.732,0.958 \end{tabular} &} \begin{tabular}[x]{@{}c@{}}38\\$86.8\%$ \\ 0.719,0.956 \end{tabular} & \begin{tabular}[x]{@{}c@{}}34\\$79.4\%$ \\ 0.621,0.913 \end{tabular} & \begin{tabular}[x]{@{}c@{}}31\\$54.8\%$ \\ 0.360,0.727 \end{tabular} & N/A \\
\hline
t\_\HeavyStart
  & 100  & \cut{\begin{tabular}[x]{@{}c@{}}71\\$88.7\%$ \\ 0.790,0.950 \end{tabular}   &  \begin{tabular}[x]{@{}c@{}}61\\$85.2\%$ \\ 0.738,0.930 \end{tabular}  & \begin{tabular}[x]{@{}c@{}}60\\$85\%$ \\ 0.734,0.929 \end{tabular} & \begin{tabular}[x]{@{}c@{}}59\\$86.4\%$ \\ 0.750,0.939 \end{tabular} & 
\begin{tabular}[x]{@{}c@{}}59\\$81.3\%$ \\ 0.690,0.903 \end{tabular} & \begin{tabular}[x]{@{}c@{}}58\\$79.3\%$ \\ 0.666,0.888 \end{tabular} &
} \begin{tabular}[x]{@{}c@{}}56\\$69.6\%$ \\ 0.559,0.812 \end{tabular} & \begin{tabular}[x]{@{}c@{}}55\\$56.4\%$ \\ 0.423,0.697 \end{tabular} & \begin{tabular}[x]{@{}c@{}}50\\$50\%$ \\ 0.355,0.645 \end{tabular} & N/A \\
\hline
m\_\Aggressive
 & 75  & \cut{\begin{tabular}[x]{@{}c@{}}65\\$76.9\%$ \\ 0.648,0.845 \end{tabular}  & \begin{tabular}[x]{@{}c@{}}57\\$73.7\%$ \\ 0.603,0.844 \end{tabular} & \begin{tabular}[x]{@{}c@{}}52\\$76.9\%$ \\ 0.631,0.875 \end{tabular}  & \begin{tabular}[x]{@{}c@{}}50\\$76\%$ \\ 0.618,0.869 \end{tabular}  & \begin{tabular}[x]{@{}c@{}}49\\$73.5\%$ \\ 0.589,0.851 \end{tabular}   & 
 \begin{tabular}[x]{@{}c@{}}42\\$73.8\%$ \\ 0.580,0.861 \end{tabular}  &}
 \begin{tabular}[x]{@{}c@{}}39\\$64.1\%$ \\ 0.472,0.788 \end{tabular}  & N/A & N/A & N/A  \\
\hline
m\_\HeavyStart\_2
  & 81 & \cut{ \begin{tabular}[x]{@{}c@{}}50\\$100\%$ \\ 0.929,1 \end{tabular}  & \begin{tabular}[x]{@{}c@{}}42\\$100\%$ \\ 0.916,1 \end{tabular}    & \begin{tabular}[x]{@{}c@{}}42\\$100\%$ \\ 0.916,1 \end{tabular} & \begin{tabular}[x]{@{}c@{}}41\\$100\%$ \\ 0.914,1 \end{tabular}  &\begin{tabular}[x]{@{}c@{}}38\\$100\%$ \\ 0.907,1 \end{tabular} & \begin{tabular}[x]{@{}c@{}}37\\$100\%$ \\ 0.905,1 \end{tabular}   & }\begin{tabular}[x]{@{}c@{}}36\\$100\%$ \\ 0.903,1 \end{tabular}  & \begin{tabular}[x]{@{}c@{}}33\\$100\%$ \\ 0.894,1 \end{tabular} &
\begin{tabular}[x]{@{}c@{}}30\\$90\%$ \\ 0.735,0.979 \end{tabular} & N/A \\
\hline
m\_\HeavyStart\_1
  & 86  & \cut{ \begin{tabular}[x]{@{}c@{}}64\\$100\%$ \\ 0.943,1 \end{tabular}     & \begin{tabular}[x]{@{}c@{}}52\\$100\%$ \\ 0.932,1 \end{tabular}   & \begin{tabular}[x]{@{}c@{}}49\\$100\%$ \\ 0.927,1 \end{tabular}    & \begin{tabular}[x]{@{}c@{}}49\\$100\%$ \\ 0.927,1 \end{tabular}  & \begin{tabular}[x]{@{}c@{}}47\\$100\%$ \\ 0.925,1 \end{tabular}   &  \begin{tabular}[x]{@{}c@{}}46\\$100\%$ \\ 0.923,1 \end{tabular}    & }\begin{tabular}[x]{@{}c@{}}45\\$100\%$ \\ 0.922,1 \end{tabular}   & \begin{tabular}[x]{@{}c@{}}44\\$100\%$ \\ 0.920,1 \end{tabular}  & \begin{tabular}[x]{@{}c@{}}43\\$100\%$ \\ 0.918,1 \end{tabular} & N/A  \\
\hline
m\_\VeryConservative  & 83  & \cut{\begin{tabular}[x]{@{}c@{}}72\\$86.1\%$ \\ 0.759,0.931 \end{tabular}   & \begin{tabular}[x]{@{}c@{}}61\\$85.2\%$ \\ 0.738,0.930 \end{tabular}  &  \begin{tabular}[x]{@{}c@{}}60\\$85\%$ \\ 0.734,0.929 \end{tabular}  & \begin{tabular}[x]{@{}c@{}}60\\$85\%$ \\ 0.734,0.929\end{tabular}  & \begin{tabular}[x]{@{}c@{}}58\\$84.5\%$ \\ 0.72.6,0.927 \end{tabular} & \begin{tabular}[x]{@{}c@{}}55\\$81.8\%$ \\ 0.691,0.909 \end{tabular}  &} \begin{tabular}[x]{@{}c@{}}53\\$81.1\%$ \\ 0.680,0.906 \end{tabular}   & \begin{tabular}[x]{@{}c@{}}51\\$80.4\%$ \\ 0.669,0.902 \end{tabular} & \begin{tabular}[x]{@{}c@{}}49\\$77.6\%$ \\ 0.634,0.882 \end{tabular} &\begin{tabular}[x]{@{}c@{}}44\\$81.8\%$ \\ 0.673,0.918 \end{tabular} \\
\hline
m\_\HeavierStart  & 73 & \cut{ \begin{tabular}[x]{@{}c@{}}40\\$95\%$ \\ 0.831,0.994 \end{tabular}  &  \begin{tabular}[x]{@{}c@{}}27\\$100\%$ \\ 0.872,1 \end{tabular}  & \begin{tabular}[x]{@{}c@{}}26\\$100\%$ \\ 0.868,1 \end{tabular}  & \begin{tabular}[x]{@{}c@{}}24\\$100\%$ \\ 0.858,1 \end{tabular}  & \begin{tabular}[x]{@{}c@{}}22\\$100\%$ \\ 0.846,1 \end{tabular}  & \begin{tabular}[x]{@{}c@{}}22\\$100\%$ \\ 0.846,1 \end{tabular} &} \begin{tabular}[x]{@{}c@{}}22\\$100\%$ \\ 0.846,1 \end{tabular} & \begin{tabular}[x]{@{}c@{}}22\\$100\%$ \\ 0.846,1 \end{tabular} & \begin{tabular}[x]{@{}c@{}}22\\$100\%$ \\ 0.846,1 \end{tabular} &
\begin{tabular}[x]{@{}c@{}}21\\$95.2\%$ \\ 0.762,0.999 \end{tabular} \\
\hline
\hline \begin{tabular}[x]{@{}c@{}}{\bf Rehearsal~$ i\backslash$}\\{\bf Condition}\end{tabular}
 &  &\centering  $i$ $=$ $11$ &\centering  $i$ $=$ $12$ &   &  \\
\hline
m\_\HeavierStart &   & \begin{tabular}[x]{@{}c@{}}21\\$85.7\%$ \\ 0.637,0.970 \end{tabular} & \begin{tabular}[x]{@{}c@{}}17\\$82.4\%$ \\ 0.566,0.962 \end{tabular}  & & \\
 \hline
\end{tabular}%
}
\caption{$\mathbf{NumSurvived}(i)$/$\mathbf{NumReturned}(i)$ with $95\%$ binomial confidence intervals. $m=$ ``mnemonic," $t = $``text"}
\label{tab:survived}
\end{table}

\begin{figure}[tb]
\centering
\includegraphics[scale=0.4]{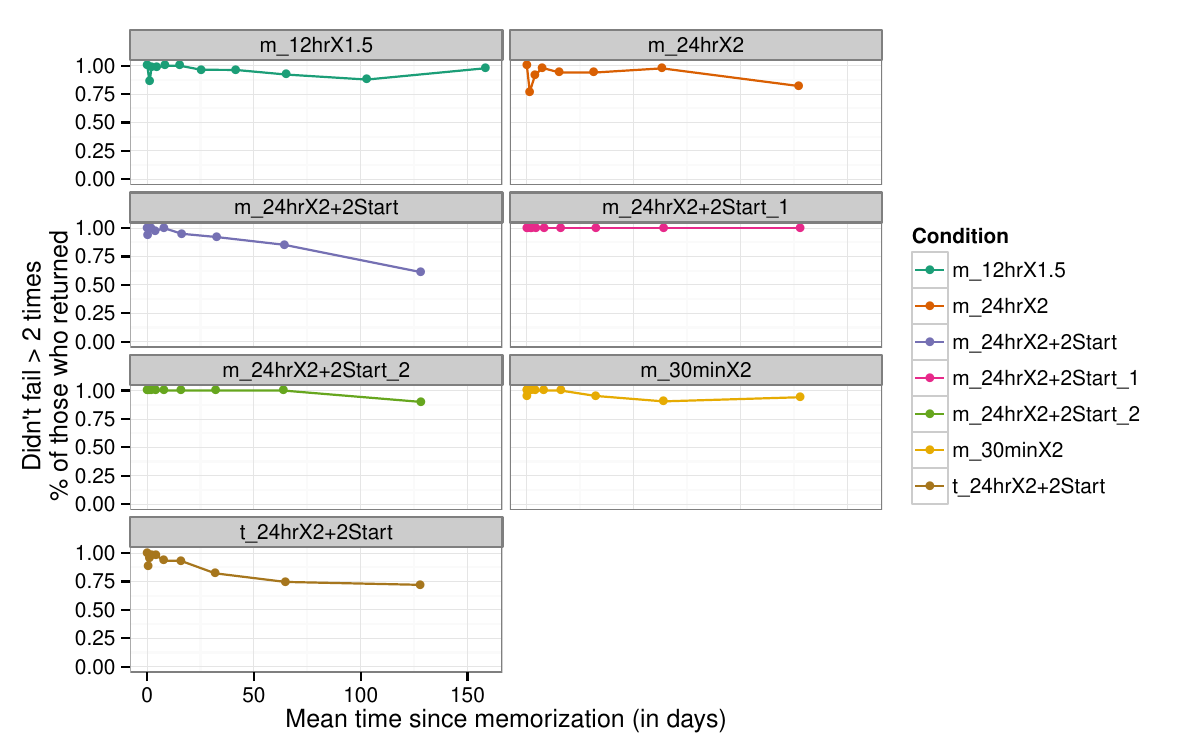}
\caption{$\mathbf{NumRemembered}\paren{i}/\mathbf{NumReturned}\paren{i}$ vs $\mathbf{Time}\paren{i}$}
\label{fig:ConditionalSuccessMethod2}
\end{figure}

\begin{figure}[tb]
\centering
\includegraphics[scale=0.4]{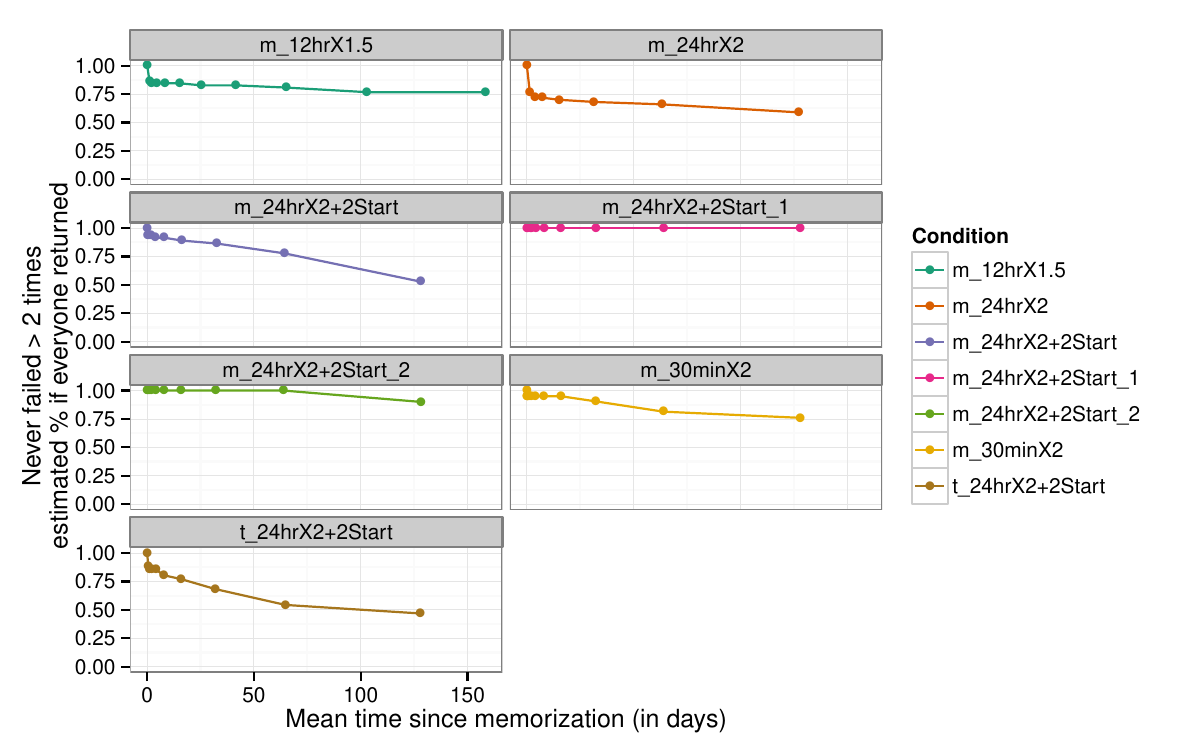}
\caption{$\mathbf{EstimatedSurvival}\paren{i}$ vs $\mathbf{Time}\paren{i}$}
\label{fig:EstimatedFacetedConditionalSurvivalProbability2}
\end{figure}

We compare three different metrics to estimate the survival rate of participants in our study under ideal circumstances (e.g., if all of our participants were always able to return to rehearse in a timely manner). Our first estimate is $\mathbf{EstimatedSurvival}\paren{i} =$ \[ \prod_{j=1}^i \frac{\mathbf{NumSurvived}\paren{j}}{\mathbf{NumSuccessfulReturned}(j)} \ . \]  We plot this value in Figures \ref{fig:EstimatedFacetedConditionalSurvivalProbability2}\cut{ and \ref{fig:EstimatedSurvivalProbability2}}. 

\begin{figure}
\centering
\includegraphics[scale=0.4]{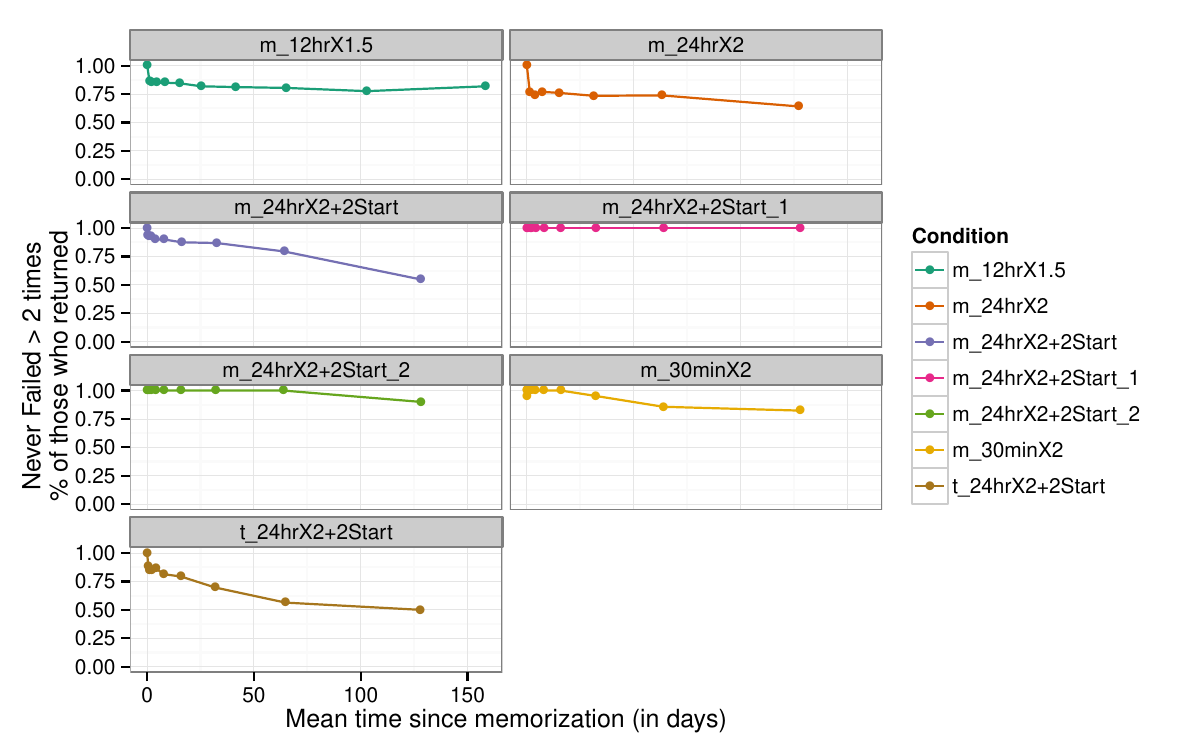}
\caption{$\mathbf{ObservedSurvival}\paren{i}$ vs $\mathbf{Time}\paren{i}$}
\label{fig:EstimatedSurvivalProbabilityMethod2}
\end{figure}

Our second estimate, shown in Figure \ref{fig:EstimatedSurvivalProbabilityMethod2}, is quite simple: \[ \mathbf{ObservedSurvival}\paren{i} = \frac{\mathbf{NumSurvived}\paren{i}}{\mathbf{NumReturned}\paren{i}} \ . \] 

Our first estimate could be biased if participants are less likely to return for future rehearsals whenever they think they might forget their words. Our second estimate could be biased if participants were less likely to return for future rehearsal rounds after previous failures. However, we did not observe any obvious correlation between prior failure and the return rate. Sometimes the return rate was higher for participants who had failed earlier than for participants who had never failed and sometimes the return rate was lower. Furthermore, in our survey of participants who did not return in time for a rehearsal round no one self-reported that they did not return because they were not confident that they would be able to remember (see Section \ref{subsubsec:SurveyResults}). Both methods consistently yielded close estimates. 

\begin{figure}
\centering
\includegraphics[scale=0.40]{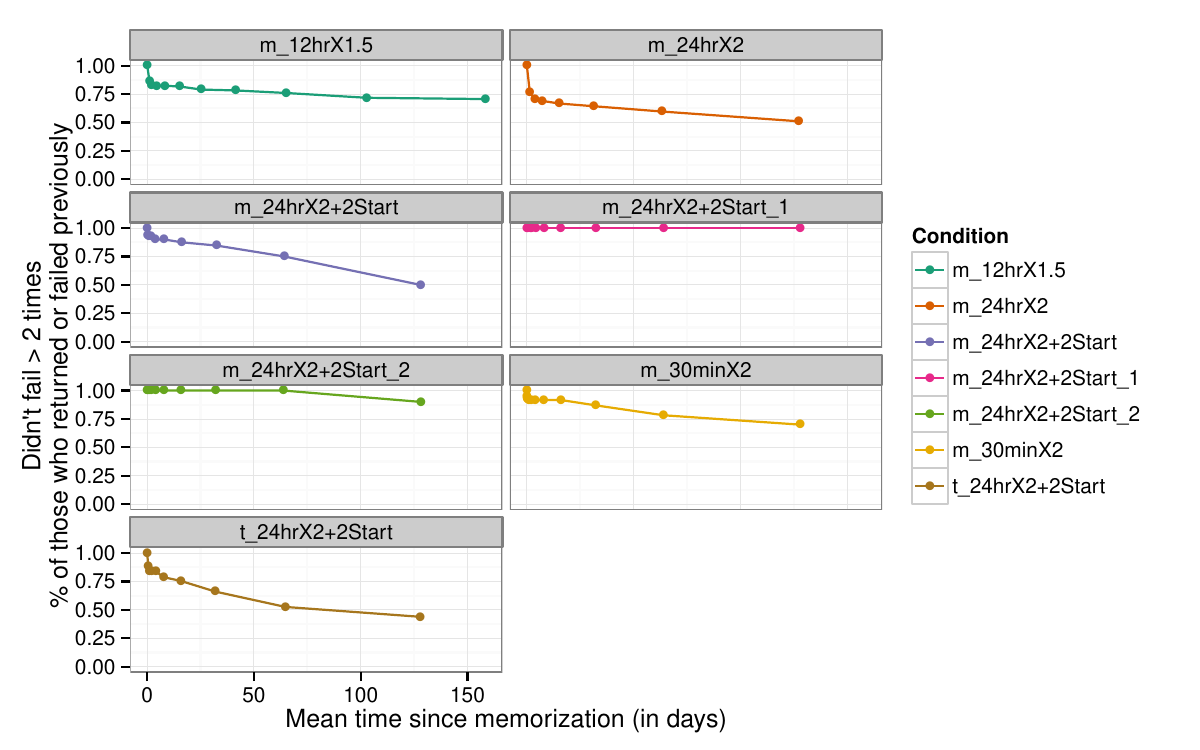}
\caption{$\mathbf{PessimisticSurvivalEstimate}\paren{i}$ vs $\mathbf{Time}\paren{i}$.}
\label{fig:FacetedTotalSurvival}
\end{figure}

We also consider a pessimistic estimate of the survival rate (see Figure \ref{fig:FacetedTotalSurvival})  
 $\mathbf{PessimisticSurvivalEstimate}\paren{i}=$
\[ \frac{\mathbf{NumSurvived}\paren{i}}{\mathbf{NumSuccessfulReturned}\paren{i}+\mathbf{NumFailed}\paren{i}} \ .\]  
Here, $\mathbf{NumFailed}\paren{i}$ counts the number of participants who failed to remember at least one of his action-object pairs with $<3$ guesses during {\em any} rehearsal $j\leq i$ --- even if that participant did not return for later rehearsal rounds. This estimate is most likely overly pessimistic because it includes every participant who failed early on during the study before dropping out, but it excludes every participant who dropped out without failing. For example, a participant who succeeded through rehearsal $5$ but could not return for rehearsal $6$ would not be included in the estimate for $i\geq 6$. However, if our participant had failed during round $2$ before not returning for rehearsal $3$ then s/he would still be included in the estimate.\cut{\footnote{As an example, consider a participant who correctly remembered his stories during the first three rehearsals, but was not able to return for the fourth rehearsal (e.g., because he went on vacation). The results of this participant would be dropped. However, if the same participant had failed during round three then his results would be included because we would not have asked him to return for the fourth rehearsal while he was on vacation. Suppose that participants who return for rehearsal one succeed with probability $0.99$ and that any participant who is able to return for rehearsal $i > 1$ will succeed with probability $1$. The true survival rate would be $99\%$ for all $i>0$, but if each participant is not able to return to complete rehearsal $i$ independently with probability $p > 0$ then the total survival rate among participants who always returned will always tend to $0$ as $i$ grows. }.}

\subsection{Survey Results} \label{subsubsec:SurveyResults} 
\begin{figure}[htb]
\begin{center}
\hspace*{-0.1\columnwidth}
\includegraphics[width=1.1\columnwidth]{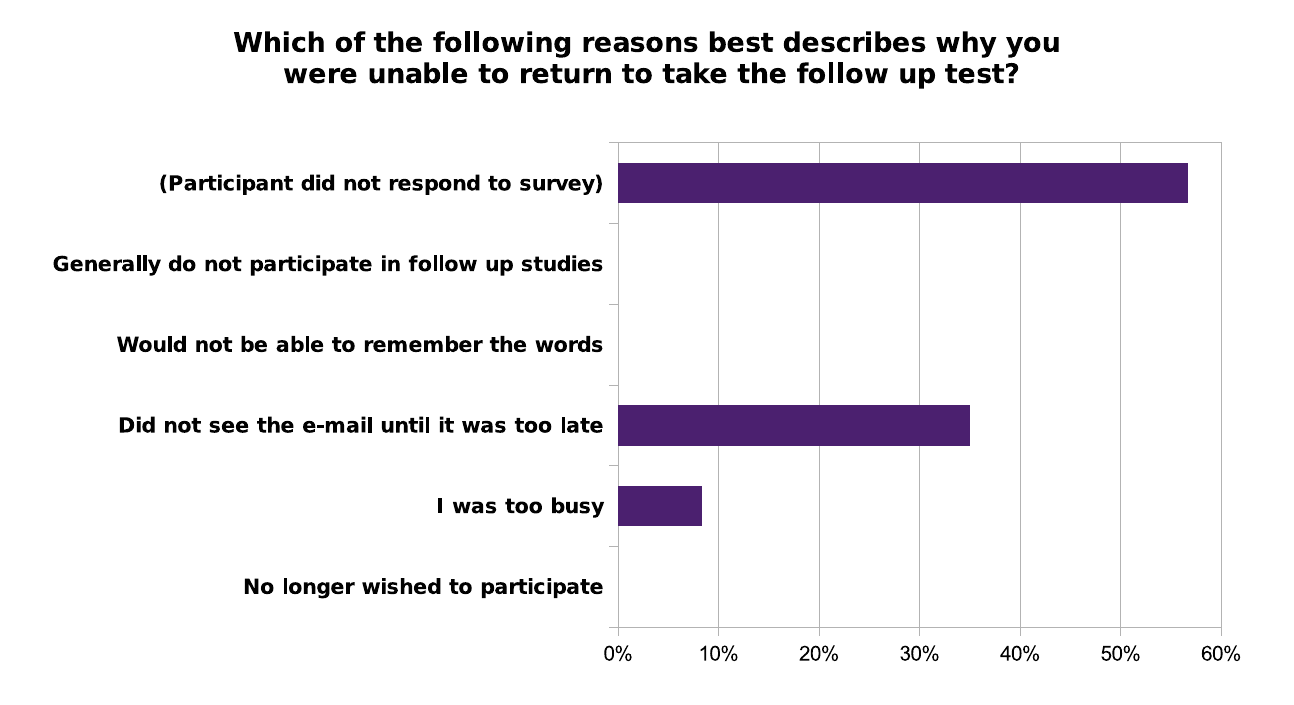}
\end{center}
%
%
%
%
%
\caption[Survey: Which of the following reasons best describes why you were unable to return to take the follow up test?]{Survey Results}
\label{fig:SurveyReturned}
\end{figure}

We surveyed $61$ participants who did not return to complete their first rehearsal to ask them why they were not able to return. The results from our survey are presented in Figure \ref{fig:SurveyReturned}. The results from our survey strongly support our hypothesis that the primary reason that participants do not return is because they were too busy, because they did not get our follow up message in time, or because they were not interested in interacting with us outside of the initial Mechanical Turk task, and not because they were convinced that they would not remember the story --- no participant indicated that they did not return because they thought that they would not be able to remember the action-object pairs that they memorized.

\paragraph*{Fun} We had several participants e-mail us to tell us how much fun they were having memorizing person-action-object stories. The results from our survey are also consistent with the hypothesis that memorizing person-action-object stories is fun (e.g., no participants said that they no longer wished to participate in the study).

\subsection{Statistical methods} \label{sec:CoxRegression}

We used Cox regression to perform survival analysis, and compare different study conditions. Survival analysis relates the time that passes before a failure event (e.g., a user forgets one of his action-object pairs during a rehearsal) to covariates (e.g., mnemonic/text, rehearsal schedule, number of words memorized) that may be associated with this quantity of time. Cox regression is an appropriate tool for our study because it can deal with participants who dropped out of the study before they failed (e.g., participants who did not return for a rehearsal in a timely manner). 

To run Cox regression we need to have the following data for each participant: whether or not they failed, the time of failure for participants who did fail and the time a participant dropped out of the study for participants who were dropped before their first failure. The output of Cox regression is a set of regression coefficients $\beta_1,\ldots,\beta_k$, which tell us how different study conditions affect the survival rate compared to a baseline condition\footnote{We use the proportional hazard model to compare the risk of failure for participants in different study conditions. In particular, given a baseline study condition we let  $\lambda_0(t)$ denote the underlying hazard function --- a function describing the risk that a participant fails to remember his action-object pairs at time $t$. Given explanatory variables $x_1,\ldots,x_p$ we use the function 
$\lambda\paren{t~\vline~x_1,\ldots,x_p} = \lambda_0\paren{t} \exp\paren{\beta_1x_1 + \ldots + \beta_p x_p}$,
to compare the risk for participants in different study conditions and we use Cox regression to compute the regression coefficients $\beta_1,\ldots,\beta_p$ for each explanatory variable.}. For example, suppose that our baseline condition was the t\_\HeavyStart{} condition then the regression coefficient $\beta_m$ would tell us how the survival rate in the m\_\HeavyStart~condition compared to the survival rate in the t\_\HeavyStart{} condition. If $\exp\paren{\beta_m} < 1$ (resp. $\exp\paren{\paren{\beta_m}} > 1$) then participants in the mnemonic condition are less likely (resp. more likely) to fail at any given time than participants in the baseline text condition.

\subsection{Findings} \label{subsec:findings}

\subsubsection{Reliable Recall of Multiple Passwords} We found that $ 82\%$ of participants in the m\_\VeryConservative~condition who were always able to return were  able to remember all four of their secret action-object pairs during all ten of their rehearsal rounds over a time period of $158$ days. This value ($\mathbf{NumSurvived}\paren{10}/\mathbf{NumReturned}\paren{10} = 0.818$) closely matches our estimated survival rate  $\mathbf{EstimatedSurival}\paren{10} = 0.765$. As explained before, eight secrets (four actions and four objects) could be used to form $14$ different passwords using the Shared Cues password management scheme\cite{blockiNaturallyRehearsingPasswords}. As we can see from Table \ref{tab:remembered}, most of the participants who did not survive forgot their action-object pairs during the first rehearsal. 
 
\subsubsection{Reliable Recall of a Single Password} $100\%$ of participants who were asked to memorize one or two PAO stories (e.g., one password) were able to remember all of their secret action-object pairs during their first eight rehearsals over a time period of $63.6$ days. $100\%$ of participants in the m\_\HeavyStart\_1 condition also remembered their PAO story during the final rehearsal on day $128$. The $95\%$ confidence interval for the fraction of participants in the m\_\HeavyStart\_1 (resp. m\_\HeavyStart\_2) condition who always remember their action-object pairs over $128$ days is $[0.918,1]$ (resp. $[0.673,0.918]$). 

\subsubsection{Effect of Rehearsal} The m\_\VeryConservative{} condition had the best survival rate. We used Cox regression to compare the survival rate for participants in the m\_\Aggressive,  m\_\VeryConservative, m\_\HeavyStart{}, and m\_\HeavierStart{} study conditions, using the m\_\Aggressive{} condition as our baseline. Our results are shown in Table \ref{tab:multipleSchedulesHypothesis}. For all three conditions m\_\VeryConservative, m\_\HeavyStart{}, and m\_\HeavierStart{} we have $\exp\paren{\beta_i} < 1$ meaning that our model predicts that participants in the baseline condition (m\_\Aggressive) were less likely to survive at any given point in time. However, the evidence for the hypothesis $\exp\paren{\beta_i} < 1$ was only statistically significant (at the $\alpha =0.05$ level) for the m\_\VeryConservative{} conditions because the confidence interval for $\exp\paren{\beta_i}$ does not contain the value $1$. A larger study would need to be done to establish statistical significance for the \HeavyStart{} or  \HeavierStart{}  conditions\footnote{Because the \HeavierStart{} condition had many rehearsals on day 1 our time window for participants to return for each of these rehearsals was more narrow than in other conditions. As a result many participants in this condition were dropped after day 1 because they were not able to return in a timely manner.}.

\begin{table}[t]
\centering
\resizebox{\columnwidth}{!}{%
\begin{tabular}{|l|c|c|c|c|c|}
\hline
Condition $x_i$ & $\beta_i$ & $\exp\paren{\beta_i}$ & $\exp\paren{-\beta_i}$ & \multicolumn{2}{p{2.2cm}|}{ \begin{tabular}[x]{@{}cc@{}} \multicolumn{2}{c}{\begin{tabular}[x]{@{}cc@{}} $95\%$ Confidence \\ Interval  for $\exp\paren{\beta_i}$ \end{tabular}}\\Lower & Upper\end{tabular}}  \\
\hline
m\_\HeavierStart & -0.8323 & 0.4351 & 2.299 & 0.1760 & 1.0752 \\
\hline
m\_\HeavyStart & -0.2610 & 0.7703 & 1.298 & 0.4108 & 1.442  \\
\hline
m\_\VeryConservative & -0.8677 & 0.4199 & 2.381 & 0.2164 & 0.8147 * \\
\hline
\end{tabular}%
}

{\footnotesize \raggedright \vspace{0.25em}

$n = 228$, number of failure events $k= 56$.

* indicates $\exp\paren{\beta_i}$ is significantly different from $1$ at the $\alpha=0.05$ level.
}

\caption[foo]{Cox regression with baseline: m\_\Aggressive{}}
\label{tab:multipleSchedulesHypothesis}
\end{table}

We contrast the results of Cox regression, which operates over the full duration of our study, with simple $t$-tests performed on estimated survival rates between conditions at a specific point in time, typically $64$ days after memorization.  We do this to examine estimated performance after passwords have been used for an extended period of time. In particular, we used a one-tailed $t$-tests for the hypotheses that $\mathbf{ObservedSurvival}\paren{C_1,i_1} >$ $\mathbf{ObservedSurvival}\paren{C_2,i_2}$ 
for $C_1=$ m\_\Aggressive{} and $C_2 \in \{ $ m\_\VeryConservative, m\_\HeavierStart, m\_\HeavyStart{} $ \}$.  We adjusted the values of $i_1$ and $i_2$ to compare the survival rates over similar time periods for participants who followed different rehearsal schedules. In particular, every rehearsal schedule had one rehearsal near day $64$ (e.g., the sixth rehearsal in the \Aggressive{} schedule was on day $63$ and the eighth rehearsal for the \VeryConservative{} schedule was on day $64.65$). Table \ref{tab:GreaterSurvival} shows these results. While the survival rates were lowest in the m\_\Aggressive{} condition the $t$-test results were not statistically significant  at the $\alpha =0.05$ level.

\subsubsection{Mnemonic vs Text Conditions} We found that participants who used the PAO mnemonic technique significantly outperform users who didn't in recalling their action-object pairs in the short term. In particular, we used a one-tailed $t$-test and tested the hypothesis 
$\mathbf{ObservedSurvival}\paren{C_1,i_1} > \mathbf{ObservedSurvival}\paren{C_2,i_2}$ for $C_1=$ m\_\HeavyStart~and $C_2 =$  t\_\HeavyStart.  We compared our conditions at two points in time: $63.6$ days after memorization $(i_1=i_2 = 8)$ and after the final rehearsal $127.6$ days after memorization $(i_1=i_2=9)$. Table \ref{tab:GreaterSurvival} shows these results. 

The evidence that participants in  m\_\HeavyStart{} perform better than participants in t\_\HeavyStart{} for the first $63.6$ days after memorization is statistically significant ($p=0.010$). However, after the final rehearsal $(i_1=i_2=9)$ the evidence for our hypothesis is no longer statistically significant. Surprisingly, during the last rehearsal round on day $127.6$, participants in the  t\_\HeavyStart{} condition were more likely to remember their action-object pairs than participants in the m\_\HeavyStart{} condition. \cut{(In fact, the $95\%$ confidence intervals in Table \ref{tab:remembered} do not intersect).} We also used Cox regression to compare the survival rates for the t$\_$\HeavyStart~and m$\_$\HeavyStart~conditions using the m$\_$\HeavyStart~ condition as a baseline. Our results are shown in Table \ref{tab:mnemonicVsText}. We have $\exp\paren{\beta_i} > 1$ for the t$\_$\HeavyStart~condition, which indicates that participants did benefit from adopting mnemonic techniques to memorize their action-object pairs. However, the hypothesis $\exp\paren{\beta_1} > 1$ (e.g., the survival rate is worse in the text condition) does not reach the level of statistical significance at the $\alpha =0.05$ level. Table \ref{tab:mnemonicVsText2} shows the results of Cox regression if we only include data from the first eight rehearsal rounds (through day $63.6$). In this case the hypothesis $\exp\paren{\beta_i} > 1$ is statistically significant.

\begin{table}[t]
\centering
\resizebox{\columnwidth}{!}{%
\begin{tabular}{|c|c|c|c|c|c|}
\hline
Condition $x_i$ & $\beta_i$ & $\exp\paren{\beta_i}$ & $\exp\paren{-\beta_i}$ & \multicolumn{2}{p{2.2cm}|}{ \begin{tabular}[x]{@{}cc@{}} \multicolumn{2}{c}{\begin{tabular}[x]{@{}cc@{}} $95\%$ Confidence \\ Interval  for $\exp\paren{\beta_i}$ \end{tabular}}\\Lower & Upper\end{tabular}}  \\
\hline
t$\_$\HeavyStart & 0.4183 & 1.519 & 0.6582 & 0.843 & 2.738  \\
\hline
\end{tabular}%
}

{\footnotesize \raggedright \vspace{0.25em}

$n = 122$, number of failure events $k= 49$.

}

\caption[foo]{Baseline: m$\_$\HeavyStart{}}
\label{tab:mnemonicVsText}
\end{table}

\begin{table}[t]
\centering
\resizebox{\columnwidth}{!}{%
\begin{tabular}{|c|c|c|c|c|c|}
\hline
Condition $x_i$ & $\beta_i$ & $\exp\paren{\beta_i}$ & $\exp\paren{-\beta_i}$ & \multicolumn{2}{p{2.2cm}|}{ \begin{tabular}[x]{@{}cc@{}} \multicolumn{2}{c}{\begin{tabular}[x]{@{}cc@{}} $95\%$ Confidence \\ Interval  for $\exp\paren{\beta_i}$ \end{tabular}}\\Lower & Upper\end{tabular}}  \\
\hline
t$\_$\HeavyStart & 1.04 & 2.829 & 0.3434 & 1.287 & 6.219 * \\
\hline
\end{tabular}%
}

{\footnotesize \raggedright \vspace{0.25em}

$n = 122$, number of failure events $k= 37$.

* indicates $\exp\paren{\beta_i}$ is significantly different from $1$ at the $\alpha=0.05$ level.
}

\caption[foo]{Baseline: m$\_$\HeavyStart{}}
\label{tab:mnemonicVsText2}
\end{table}

\begin{table}[t]
\resizebox{\columnwidth}{!}{%
\begin{tabular}{|c|c|c|c|c|}
\hline
Condition $C_1$ & $i_1$ (Day) & Condition $C_2$ & $i_2$ (Day) & $p$-value \\
\hline
m\_\HeavyStart & 8 ($63.6$) & t\_\HeavyStart & 8 ($63.6$) & $0.010$ * \\
\hline
m\_\HeavyStart & $9$ ($127.6$) & t\_\HeavyStart & $9$ ($127.6$) & $0.338$ \\
\hline
m\_\HeavyStart\_1 &  8 ($63.6$)  & \begin{tabular}[x]{@{}cc@{}} m\_\HeavyStart \\ Remark 1. \end{tabular} & 8 ($63.6$) & 0.042 *\\
\hline
m\_\HeavyStart\_2 &  8 ($63.6$)  & \begin{tabular}[x]{@{}cc@{}} m\_\HeavyStart \\ Remark 2.  \end{tabular} & 8 ($63.6$) & 0.006 * \\
\hline
m\_\HeavyStart\_1 &  8 ($63.6$)  & \begin{tabular}[x]{@{}cc@{}} m\_\HeavyStart \\ Remark 3.  \end{tabular} & 8 ($63.6$) & 0.0011 * \\
\hline
m\_\HeavyStart\_2 &  8 ($63.6$)  & \begin{tabular}[x]{@{}cc@{}} m\_\HeavyStart \\ Remark 3.  \end{tabular} & 8 ($63.6$) & 0.0011 * \\
\hline
m\_\HeavyStart\_1 &  8 ($63.6$)  & \begin{tabular}[x]{@{}cc@{}} m\_\HeavyStart \\ Remarks 1 and 3. \\   \end{tabular} & 8 ($63.6$) & 0.03 * \\
\hline
m\_\HeavyStart\_2 &  8 ($63.6$)  & \begin{tabular}[x]{@{}cc@{}} m\_\HeavyStart \\ Remarks 2 and 3. \end{tabular} & 8 ($63.6$) & 0.046 * \\
\hline

m\_\HeavyStart\_1 &  9 ($127.6$)  & \begin{tabular}[x]{@{}cc@{}} m\_\HeavyStart \\ Remark 1. \end{tabular} & 9 ($127.6$) & 0.011 *\\
\hline
m\_\HeavyStart\_2 &  9 ($127.6$)  & \begin{tabular}[x]{@{}cc@{}} m\_\HeavyStart \\ Remark 2.  \end{tabular} &  9 ($127.6$)  & 0.017 * \\
\hline
m\_\HeavyStart\_1 &  9 ($127.6$)  & \begin{tabular}[x]{@{}cc@{}} m\_\HeavyStart \\ Remarks 1 and 3. \\   \end{tabular} &  9 ($127.6$)  & 0.012 * \\
\hline
m\_\HeavyStart\_2 &  9 ($127.6$)  & \begin{tabular}[x]{@{}cc@{}} m\_\HeavyStart \\ Remarks 2 and 3. \end{tabular} &  9 ($127.6$)  & 0.331 \\
\hline
\begin{tabular}[x]{@{}cc@{}} m\_\HeavyStart   \end{tabular} & 8 ($63.6$) & m\_\Aggressive &  6 ($63$)  &  0.287 \\
\hline
 \begin{tabular}[x]{@{}cc@{}} m\_\HeavierStart \end{tabular} & 8 ($63.6$) &m\_\Aggressive &  6 ($63$)  & 0.146 \\
\hline
 \begin{tabular}[x]{@{}cc@{}} m\_\VeryConservative  \end{tabular} & 8 ($64.65$) & m\_\Aggressive &  6 ($63$)  & 0.228  \\
\hline
\end{tabular}%
}

{\footnotesize \raggedright \vspace{0.25em}

Remark 1. Count participant as surviving if s/he always remembered the first action-object pair.

Remark 2.  Count participant as surviving if s/he always remembered the first two action-object pairs.

Remark 3. If a participant dropped and never failed count them as surviving.

* indicates statistical significance at the $\alpha=0.05$ level.
}

\captionsetup{justification=centering}
\caption[foo]{One-Tailed $t$-tests for Hypotheses: $\mathbf{ObservedSurvival}\paren{C_1,i_1} > \mathbf{ObservedSurvival}\paren{C_2,i_2}$}
\label{tab:GreaterSurvival}
\end{table}

\subsubsection{Effect of Interference} We found that there is an interference effect. Participants performed better when memorizing one or two PAO stories. We used Cox regression with the m$\_$\HeavyStart$\_4$ condition as a baseline and found statistically significant evidence ($\alpha < 0.01$) that the survival rate was better in the m$\_$\HeavyStart$\_2$ condition ($\exp\left(\beta_i\right) = 6.905$). Because there were no failure events in the m$\_$\HeavyStart$\_1$ condition Cox regression did not converge for this condition ---$\beta_i$ approaches $-\infty$. We also used a one-tailed t-test to test the hypothesis $\mathbf{ObservedSurvival}\paren{C_1,i_1} > \mathbf{ObservedSurvival}\paren{C_2,i_2}$ for the conditions $C_1 \in \{$m$\_$\HeavyStart$\_1$, m$\_$\HeavyStart$\_2\}$  and $C_2 =$ m$\_$\HeavyStart$\_4$. We tested the hypothesis $127.6$ days after memorization (e.g., $i_1=i_2=9$). The results are shown in Table \ref{tab:GreaterSurvival}. The evidence for both hypotheses was statistically significant. In fact, we can even confirm much stronger versions of these hypotheses. For example, the survival rate in the m$\_$\HeavyStart$\_1$ condition is greater than the survival rate in the m$\_$\HeavyStart$\_4$ even if we only count failures on the first action-object pair and even if we treat every participant $P$ from the m$\_$\HeavyStart$\_4$ condition who dropped without failing as if they had survived.

\section{Related Work} \label{sec:UserStudyRelated}
\subsection{Spaced Repetition}

Pimsleur\cite{Pimsleur1967} proposed a rehearsal schedule to help people memorize unfamiliar vocabulary words. He suggested rehearsing after 5 seconds, 25 seconds, 2 minutes, 10 minutes, 5 hours, 1 day, 5 days, 20 days, etc.  Pimsleur based his recommendations on previous empirical studies\cite[pp.~726~ff]{woodworth1954experimental}. Blocki et al.\cite{blockiNaturallyRehearsingPasswords} developed a usability model for password management schemes based on an assumption they called the expanding rehearsal assumptions. Loosely, this assumption states that a user will be able to remember a secret by following a rehearsal schedule similar to the one proposed by Pimsleur. The application SuperMemo\cite{wozniak2007supermemo} uses a similar rehearsal schedule to help users remember flashcards. Wozniak and Gorzelanczyk conducted an empirical study to test these rehearsal schedules\cite{superMemo}. In their study undergraduate students were asked to memorize and rehearse vocabulary words for a foreign language by following a rehearsal schedule very similar to the expanding rehearsal schedule. Wozniak and Gorzelanczyk tracked each students performance with each particular vocabulary word and used that information to estimate how difficult each word was. If a word was deemed `difficult' then the length of the time interval before the next rehearsal would only increase by a small multiplicative constant (e.g., $1.5$) and if the word was judged to be `easy' then this time interval would increase by a larger multiplicative constant (e.g., $4$). 

 We  stress two key differences in our study: First, because we are asking the user to memorize secrets that will be used to form passwords our rehearsal schedule needs to be conservative enough that our user will consistently be able to remember his secrets during each rehearsal. In other studies the information participants were asked to memorize (e.g., vocabulary words) was not secret so a participant could simply look up the correct answer whenever they forgot the correct answer during a rehearsal. By contrast, in the password setting a recovery mechanism may not always be available --- users are advised against writing down passwords and organizations have been held liable for damages when they do not properly encrypt their passwords \cite{lawsuit:RockYouResult2011}. Second, in our study we ask participants to memorize secrets by following the Person-Action-Object mnemonic techniques. Because these secrets may be easier or harder to memorize than other information like vocabulary words the ideal rehearsal schedule should be tailored to particular mnemonic techniques adopted by the user. Previous studies have demonstrated that cued recall is easier than pure recall (see for example \cite{memory:textbook:baddeley1997}) and that we have a large capacity for visual memories\cite{Memory:10000Pictures:standingt1973}. However, we are not aware of any prior studies which compare cued recall and pure recall when participants are following a rehearsal schedule similar to the one suggested by the expanding rehearsal assumption. \\

\subsection{Spaced Repetition -- Applications to Passwords}
\paragraph{Password Management Schemes}
 While there are many articles, books, papers and even comics 
about selecting strong individual passwords
\cite{burnett2005perfect,XKCDhorsebatterystaplecorrect,Gaw:2006:PMS:1143120.1143127,yan2004password,timePimpMyPassword,guideline:DOD1985,guideline:NIST2009,guideline:lifehacker}, 
there has been little work on \emph{password management schemes}---systematic 
strategies to help users create and remember multiple passwords---that are both usable and secure. Bonneau et al. \cite{bonneau2012quest} evaluated several alternatives to text passwords (e.g., graphical passwords, password management software, single-sign-on, federated authentication) finding that, while each alternative had its advantages, none of the alternatives were strictly better than text passwords. Florencio et al. \cite{florenciopassword} argued that any usable password management scheme\footnote{They use the term ``password portfolios."} cannot require users to memorize unique random passwords. They suggested that users adopt a tiered password management scheme with a unique password for high, medium and low security accounts. Blocki et al. \cite{blockiNaturallyRehearsingPasswords} recently proposed designing password management schemes that maximized the natural rehearsal rate for each of the secrets that the user had to memorize subject to minimum security constraints. Our study is heavily motivated by their work, which we already described in Section \ref{sec:Background}.

\paragraph{Slowly Expanding Password Strength} Bonneau and Schechter conducted a user study in which participants were encouraged to slowly memorize a strong $56$ bit password using spaced repetition\cite{BS14}. Each time a participant returned to complete a distractor task he was asked to login by entering his password. During the first login the participant was shown four additional random characters and asked to type them in after his password. To encourage participants to memorize these four characters they would intentionally wait a few seconds before displaying them to the user the next time he was asked to login to complete a distractor task. Once a participant was able to login several times in a row (without waiting for the characters to be displayed) they would encourage that participant to memorize four additional random characters in the same way. They found that $88\%$ of participants were able to recall their entire password without any prompting three days after the study was completed. 

There are several key difference between their study and ours: First, in our study participants were asked to memorize their entire password at the start of the study. By contrast, Bonneau and Schechter encouraged participants to slowly memorize their passwords.  Second, Bonneau and Schechter did not tell participants that their goal was to slowly memorize a strong $56$ bit password --- users were led to believe that the distractor task was the purpose of the study. By contrast, in our study we explicitly told participants that their goal was to remember their words (without writing them down). Finally, participants in our study were given fewer chances to rehearse their passwords and were asked to remember their passwords over a longer duration of time (4 months vs 2 weeks). Bonneau and Schechter asked participants to login $90$ times over a two week period. In our study participants were asked to rehearse {\em at most} $12$ times over a period of $127+$ days. We believe that the results of our study could be used to help improve the password strengthening mechanism of Bonneau and Schechter --- see discussion in Section \ref{sec:UserStudyDiscussion}.

\subsection{System Assigned Passwords. }
Empirical studies have shown that many user-selected passwords are easily guessable \cite{bonneau2012science}. A user study conducted by Shay et al. \cite{usabilitystudy:xkcd} compared several different methods of generating system assigned passwords for users to memorize (e.g., three to four random words, $5$ or $6$ random characters). They found that users had difficulty remembering system assigned passwords  48--120 hours after they had memorized it. In fact, users had more difficulty when asked to memorize three to four random words from a small dictionary than when they were asked to remember $5$ to $6$ random characters. Participants in their study were not asked to follow any particular mnemonic techniques, and were not asked to follow a rehearsal schedule.  

\subsection{Password Composition Policies} 
Another line of work on passwords has focused on composition policies (policies which restrict the space passwords that users can choose)\cite{usability:compositionPolicies,yan2004password,blockiPasswordComposition,proctor2002improving}. These policies may negatively effect usability (e.g., users report that their passwords are more difficult to remember\cite{usability:compositionPolicies,yan2004password}) and also have adverse security effects (e.g., users are more likely to write down their passwords\cite{usability:compositionPolicies,proctor2002improving}, some restrictive composition policies can actually result in a weaker password distribution\cite{blockiPasswordComposition,usability:compositionPolicies}). 

\section{Discussion} \label{sec:UserStudyDiscussion}

\paragraph*{Password Expiration Policies} Following NIST guidelines\cite{guideline2006nist} many organizations require users to change their passwords after certain period of time (e.g., thirty days). The desired behavior is for users to select a random new password that is uncorrelated with their previous passwords. We contend that these policies adversely affect usability and security. Memorizing a new password requires effort and users are typically only willing to invest a limited amount time and energy memorizing new passwords. Our experiments indicate that most of the effort to memorize and rehearse a password is spent in the first week after the new password is chosen. By forcing users to reset their password frequently an organization forces its users to remain within the most difficult rehearsal region. There is strong empirical evidence that users respond to password expiration policies by selecting weak passwords and/or selecting new passwords that are highly correlated with one of their old passwords (e.g., old\_password+$i$ for $i = 1,2,\ldots$) effectively canceling out any security gains\cite{zhang2010security}. We contend that a more productive policy would ask participants to slowly strengthen their passwords over time using spaced repetition (see discussion below).

\paragraph*{Strengthening Passwords Over Time} Our results suggest that the password strengthening mechanism of Bonneau and Schechter \cite{BS14} could be improved by adopting the PAO story mnemonic used in our study and by using a rehearsal schedule like \HeavyStart~to help predict when a user has memorized his new secret. Recall that in their mechanism a user authenticates by typing in his old password and then by typing a random character or word that is displayed next to the password box. To encourage participants to memorize this secret character/word, the user will not be shown the random character/word for several seconds, allowing a user who has memorized the secret to authenticate faster than one who has not. At some point the mechanism will predict that the user has memorized his new secret. At this point this secret is permanently appended to the user\rq{}s password so that the user must remember this secret to authenticate --- he can no longer wait for the character/word to be displayed. 

We remark that, instead of requiring the user to memorize a new random character/word to append to his password, it may be easier for the user to memorize a random action-object pair using the PAO mnemonic techniques from this study. Participants in the mnemonic$\_$\HeavyStart$\_1$ condition remembered their secret action-object pairs perfectly through $128$ days with only $9$ rehearsals. We also remark that the \HeavyStart~rehearsal schedule could provide a reasonable basis for predicting when a user has memorized his new action-object pair. In particular, a rehearsal schedule could help us predict how long the user will be able to remember his new action-object pairs without rehearsing again. If it is safe to assume that the user will return to authenticate before this point then we would argue that it is safe to predict that the user has memorized his secret action-object pair.

 Another interesting observation from our study was that participants in the m\_\Aggressive\_4 condition who remembered their action-object pairs during the first two rehearsal on days 1 and 3 were actually more likely to survive through rehearsal 6 (on day $63$) than participants in the m\_\HeavyStart\_4 condition who remembered their action-object pairs through rehearsal 4 (on day $3.6$) were to survive through the corresponding rehearsal 8 (on day $63.6$) --- though this result was not significant at the $p=0.05$ level. We hypothesize that a user's ability to remember a particular set of action-object pairs after a challenging rehearsal interval (e.g., only $77\%$ of participants in the m\_\Aggressive\_4~condition who returned for the first rehearsal on day 1 remembered their action-object stories) is better indicator of that user's future success for those particular action-object pairs than performance on less challenging rehearsal intervals. This hypothesis could also help us to predict when a user has memorized a new action-object pair. However, more studies are necessary to properly test this hypothesis.

\paragraph*{Mitigating Initial Forgetting}
We found that much of the forgetting in our study occurred during the first test period. This finding leads us to suggest three mechanisms to help ensure that users will remember their action-object pairs in the Shared Cues password management scheme: 1) Start with a shorter initial time gap between the memorization phase and the first rehearsal (e.g., 3 hours or 6 hours). 2) Instruct the user to wait $12$ hours after he has memorized the PAO stories before using the secret action-object pairs to form passwords. If the user can still remember his PAO stories after $12$ hours then he can go ahead and use those stories to create passwords. 3) Implement a temporary recovery mechanism which allows a user who can remember one or two of his action-object pairs to recover his other action-object pairs during the first 24 hours (e.g., $98.6\%$ of participants in the  m\_\VeryConservative~condition  remembered their first action-object pair after $12$ hours).

\paragraph*{Natural Rehearsals} The usability model of Blocki et al.\cite{blockiNaturallyRehearsingPasswords} was based on an assumption about human memory. Formally, their expanding rehearsal assumption says that a user will continue to remember a secret $s$ if he rehearses it at least once during each of the time intervals $\left[t_0, t_1 \right),\left[t_1,t_2\right),\ldots$ where $t_0=0$ and the length of the $i$'th interval is $t_{i+1}-t_i = b\AssociationStrength{s}^i$. Here, $\AssociationStrength{s} > 1$ is a constant which may depend on the strength of the mnemonic techniques used to memorize the secret $s$, and $b$ is a base unit of time. Observe that the length of these  rehearsal intervals grows exponentially with the number of prior rehearsals $(i)$. Our user study provides evidence that users can remember $4$ PAO stories following the \VeryConservative~rehearsal schedule (e.g., $b=12$ hours, $\AssociationStrength{s}=1.5$). 

If the user needs to recall the secret $s$ in the course of a normal login during the time interval $\left[t_i,t_{i+1}\right)$ then we say that this rehearsal requirement was satisfied naturally. If the user does not rehearse $s$ naturally during the interval $\left[t_i, t_{i+1} \right)$ then he would need to be reminded to do an extra rehearsal to ensure that he does not forget $s$. Blocki et al.\cite{blockiNaturallyRehearsingPasswords} suggested that we quantify the usability of a password management scheme by predicting how many extra rehearsals ($XR_{\infty}$) the user would need to perform over his lifetime to remember all of his password related secrets. The value on $XR_{\infty}$ will depend on how frequently the user logs into each of his accounts (in additions to the parameters $b$ and $\AssociationStrength{}$). 

We previously observed that a user could form $14$ passwords from $4$ PAO stories by adopting the Shared Cues scheme of Blocki et al.\cite{blockiNaturallyRehearsingPasswords}. Following Blocki et al. \cite{blockiNaturallyRehearsingPasswords}
we provide a sense of the extra rehearsal effort necessary to remember all four PAO stories. Table \ref{tab:ExtraRehearsals} predicts how many extra rehearsals ($XR_{\infty}$) the user would need to do over his lifetime to ensure
that he remembers all $4$ of his PAO stories in expectation. We used the parameters $b=12$ hours and $\AssociationStrength{s}=1.5$ because most users in our study were able to remember $4$ PAO stories by following the \VeryConservative~rehearsal schedule. To make these predictions we also assume that we know how frequently the user visits each of his $14$ accounts on average (e.g., daily, weekly, monthly) and that the user's visitation schedule is well-modeled by a Poisson arrival process. We consider three types of user profiles (Active, Typical and Infrequent).   

 The predictions in Table \ref{tab:ExtraRehearsals} indicate that Active and Typical users could maintain $14$ secure passwords with minimal rehearsal effort. Table \ref{tab:ExtraRehearsals}  also predicts how many extra rehearsals the user would need to do after $1.75$ days (the column labeled $\mathbf{E}[XR_{\infty}]$ - $\mathbf{E}[XR_{1.75}]$). We observe that most of the extra rehearsal effort is concentrated in first few days. After $1.75$ days our Active and Typical users would most likely not need to do any extra rehearsals over his lifetime to remember all $4$ PAO stories.

\begin{table}
\resizebox{\columnwidth}{!}{%
\begin{tabular}{|c|c|c|c|c|c|}
    \hline
      & \multicolumn{3}{c|}{ Number of Accounts Visited} & & \\
    {\bf User} &  Daily &  Weekly &  Monthly & $\mathbf{E}[XR_{\infty}]$ & $\mathbf{E}[XR_{\infty}-XR_{1.75}]$ \\
    \hline
    Active & 5 & 5 & 4 & 3.29 & $0.01$  \\
    \hline
    Typical & 2 & 8 & 4 & 7.81 & 0.14 \\
    \hline
    Infrequent & 0 & 2 & 12 & 30.41 & 7.41  \\
    \hline
\end{tabular}%
}
    \caption{$\mathbf{E}[XR_{\infty}]$ --- Expected number of extra rehearsals to remember $14$ passwords with $4$ PAO stories with $b=0.5$ days   and $\AssociationStrength{s}=1.5$}
    \label{tab:ExtraRehearsals}
\end{table}

\paragraph*{Mitigating the Interference Effect} One potential downside of the Shared Cues password management scheme \cite{blockiNaturallyRehearsingPasswords} is that the more secure versions of the scheme may require users to memorize multiple stories at once. For example, Blocki et al. \cite{blockiNaturallyRehearsingPasswords} suggested that users memorize $43$ stories to create $110$ unique passwords with a $\paren{43,4,1}$-sharing set family. While this scheme provides very strong security guarantees (e.g., an  adversary who has already seen one or two of the user's passwords could not break any of the user's other passwords even in an offline attack), the user would need to memorize at least $36$ of these stories just to form the first $9$ passwords. We observed an interference effect in our study suggesting that users would find it difficult to memorize so many stories at once. It is likely that the interference effect is, at least partially, due to user fatigue (e.g., participants who memorized four PAO stories had less mental energy to expend memorizing each action-object pair than participants who only memorized one or two PAO stories). One potential way to mitigate the interference effect would be have user\rq{}s follow a staggered schedule in which they memorize two new PAO stories at a time. Further studies are needed to test this hypothesis. Another important research problem is to construct $\paren{n,\ell,\gamma}$-sharing set families that expand gracefully so that the user does not need to memorize too many stories at the same time  (e.g., for every $t$ we seek to minimize the number of action-object pairs that a user would need to memorize to form the first $t$ passwords).

\section*{Acknowledgments}
This work was supported by the NSF Science and Technology TRUST, the AFOSR MURI on Science of Cybersecurity, CUPS IGERT grant DGE-0903659 and NSF grant CNS1116776. 

\bibliographystyle{IEEEtran}
\bibliography{password}


\appendix
\subsection{List of People, Actions and Objects from the User Study} \label{subsec:ListOfPAO}

Here are a list of the people, actions and objects we used in the study. The lists contain $92$ actions and $96$ objects respectively.

\noindent{\bf People:} Ben Afleck, Beyonce, Joe Biden, Kobe Bryant, George W Bush, Bill Clinton, Hillary Clinton, Albert Einstein, Jimmy Fallon, Pope Francis, Frodo, Gandalf, Bill Gates, Adolf Hitler,  Lebron James, Steve Jobs, Angelina Jolie,  Michael Jordan,  Nelson Mandela,  Barack Obama, Rand Paul, Ron Paul, Michael Phelps, Brad Pitt, Bart Simpson, Homer Simpson, Luke Skywalker, Justin Timberlake, Kim Jong Un, Darth Vader, Oprah Winfrey, Tiger Woods, Jay Z, Mark Zuckerberg

\noindent{\bf Actions:} aiming, aligning,  batting,  bowing, bribing, burying, canning, chipping, choking, climbing,  coating, combing, concealing,  cooking,  copying,  destroying,   dodging, drying, dueling,  egging, elbowing,  fanning,  firing, fishing,  flying,  following,  fuming, giving, gluing, gnawing,   high fiving, howling, hunting,  inhaling, judging, juicing,  jumping, kicking, kissing,  knifing,  lassoing,  leashing,  muddying, miming, marrying, mauling, mashing, mugging, moving, mopping, mowing, nipping,  nosing,  numbing, oiling, paddling, plowing, popping, puking, pulling,  punching, racing, raking, reaching, reeling, riding, rolling, rowing, saving, searing, seizing, sheering, shining,  signing, sipping, smelling, stewing, stretching, sucking, swallowing, swimming, taping, tasting, tattooing, tazing, tugging, voting, waking, waving,  weeping, welding

\noindent{\bf Objects:} ammo, ant, apple, arrow, beehive, bike, boa, boar, bomb, bunny, bus, bush, cab, cake, canoe, cat, chain, chainsaw, cheese, cheetah, chili, chime, coffee, couch, cow, daisy, dime, dish, ditch, dove, duck, fish, foot, goose, gyro, hammer, hen, home, hoof, horse, igloo, iron, jeep, jet, key, kite, leach, leaf, lime, lion, lock, mail, 
menu, microphone, moon, moose, mummy, nail, navy, nose, onion, owl, patty, phone, pill, pin, piranha, puppy, ram, rat, razor-blade, rib, roach, safe, sauce, saw, seal, shark, shoe, shoe, snake, snow, soap, sock, suit, sumo, teacup, tepee, tiger, tire, toad, toe, vase, waffle, wagon, wiener

\end{document}